\documentclass[11pt,fleqn]{article}
\usepackage{amsmath,amssymb,graphicx,epsfig,bbm}

\def \lsim{\mathrel{\vcenter
{\hbox{$<$}\nointerlineskip\hbox{$\sim$}}}}
\def \gsim{\mathrel{\vcenter
{\hbox{$>$}\nointerlineskip\hbox{$\sim$}}}}

\def\inv{\text{\small{-}}\hspace{-0.5pt}1}

\def\bea{\begin{eqnarray}}
\def\eea{\end{eqnarray}}
\def\be{\begin{equation}}
\def\ee{\end{equation}}
\def\ba{\begin{array}}
\def\ea{\end{array}}
\def\nn{\nonumber}
\def\a{& \hspace{-11pt}}
\def\b{& \hspace{-7pt}}

\font\tenrsfs=rsfs10
\font\sevenrsfs=rsfs7
\font\fiversfs=rsfs5
\newfam\rsfsfam
\textfont\rsfsfam=\tenrsfs
\scriptfont\rsfsfam=\sevenrsfs
\scriptscriptfont\rsfsfam=\fiversfs
\def\mathscr#1{{\fam\rsfsfam\relax#1}}

\begin{document}

\thispagestyle{empty}

\begin{center}

$\;$

\vspace{1cm}

{\huge \bf Scalar geometry and masses in \\[2mm] Calabi-Yau string models}

\vspace{1.5cm}

{\Large {\bf Daniel Farquet} and {\bf Claudio~A.~Scrucca}}\\[2mm] 

\vspace{0.6cm}

{\large \em Institut de Th\'eorie des Ph\'enom\`enes Physiques\\ 
Ecole Polytechnique F\'ed\'erale de Lausanne\\ 
CH-1015 Lausanne, Switzerland\\}

\vspace{0.2cm}

\end{center}

\vspace{1cm}

\centerline{\bf \large Abstract}
\begin{quote}

We study the geometry of the scalar manifolds emerging in the no-scale sector of
K\"ahler moduli and matter fields in generic Calabi-Yau string compactifications, 
and describe its implications on scalar masses. We consider both heterotic and 
orientifold models and compare their characteristics. We start from a general 
formula for the K\"ahler potential as a function of the topological compactification 
data and study the structure of the curvature tensor. 
We then determine the conditions for the space to be symmetric and show that whenever 
this is the case the heterotic and the orientifold models give the same scalar manifold. 
We finally study the structure of scalar masses in this type of geometries, assuming
that a generic superpotential triggers spontaneous supersymmetry breaking. We
show in particular that their behavior crucially depends on the parameters controlling the 
departure of the geometry from the coset situation. We first investigate the average 
sGoldstino mass in the hidden sector and its sign, and study the implications on
vacuum metastability and the mass of the lightest scalar. We next examine the 
soft scalar masses in the visible sector and their flavor structure, and study the 
possibility of realizing a mild form of sequestering relying on a global symmetry. 

\vspace{5pt}
\end{quote}

\renewcommand{\theequation}{\thesection.\arabic{equation}}

\newpage

\setcounter{page}{1}

\section{Introduction}
\setcounter{equation}{0}

The low-energy effective action of string models with minimal supersymmetry, obtained 
by compactification on a suitable internal manifold with an appropriate gauge bundle over it, 
displays a number of generic features in the weak-coupling and large-volume regime. 
One of these properties is that there are always at least two neutral chiral multiplets 
that universally appear. These are the dilaton $S$, which is related to the string coupling, and the 
overall K\"ahler modulus $T$, which is related to the volume of the compactification manifold. 
Besides these, one of course requires the presence of some non-vanishing number $n$ of charged 
chiral multiplets $\Phi^\alpha$, for the model to possibly be viable at the phenomenological level. 
Ignoring any other field, the structure of the low-energy effective theory for these fields is then quite 
universal and fixed \cite{W}. The effective K\"ahler potential, which controls the kinetic energy, is 
dominated by a non-trivial classical contribution. Additional effects coming from possible classical 
background fluxes or quantum corrections can usually be neglected, since they represent small 
corrections to the non-vanishing classical result. One then finds the following simple expression: 
\be
K = - \log \big(S+\bar S\big) - \log \big(T + \bar T - 1/3\, \Phi^\alpha \bar \Phi^\alpha\big)^3 \,.
\label{Kuniversal}
\ee 
In this situation, the scalar manifold is thus the product of two maximally symmetric coset spaces, 
with constant sectional curvatures given by $R = - 2$ for the first factor and $R = - 2/3$ for the 
second factor: 
\be
{\cal M} = \frac {SU(1,1)}{U(1)} \times \frac {SU(1,1+n)}{U(1)\times SU(n)} \,.
\label{Muniversal}
\ee
The effective superpotential, which controls the potential energy, has instead an essentially trivial classical 
contribution. Additional effects coming from possible classical background fluxes 
(see for instance \cite{GVW,GKP,KKLT} and \cite{G} for a review) 
or non-perturbative quantum corrections like gaugino condensation \cite{NP1,NP2} may thus play a crucial role. 
One can then imagine an essentially arbitrary and generic result: 
\be
W = W\big(S,T,\Phi^\alpha\big) \,.
\ee
This gives a simplified picture of the minimal situation that might be expected in string models,
which has been extensively studied in the past.

In practice, however, generic string models involve many more neutral moduli and the situation 
correspondingly gets much more involved. In particular, while the dilaton stays on its own the overall
K\"ahler modulus gets in general replaced by a number $1+m$ of K\"ahler moduli $T^A$, out of which 
$1$ controls the overall volume of the compactification manifold and the other $m$ the sizes of 
its non-trivial cycles. In addition, there may also be other qualitatively different types of non-minimal 
moduli, like for instance complex structure or bundle moduli. But these are not universally present and 
do not necessarily mix to the universal ones or the matter fields. It makes thus some sense to completely 
ignore these in a first attempt of generalization. In such a more general situation, the low-energy effective 
theory becomes less restricted and displays a much larger spectrum of possibilities \cite{FKP}. 
The K\"ahler potential is still dominated by a classical contribution, but this is no longer completely fixed 
and rather depends on the topological data of the compactification manifold and the gauge bundle over it. 
It turns out that it takes the following general form, characterized by some real functions $N^A$ which are 
in principle arbitrary but usually quadratic, and some real function $Y$ which should be homogeneous of 
degree three in its arguments but not necessarily polynomial:
\be
K = - \log \big(S + \bar S\big) - \log Y\big(T^A \!+ \bar T^A \!-\! N^A(\Phi^\alpha,\bar \Phi^\alpha)\big) \,.
\label{Kgeneral}
\ee
In this situation, the scalar manifold is no longer completely fixed. More precisely, it is the product 
of a first factor which is the same maximally symmetric coset manifold as before, with constant 
sectional curvature $R=-2$, and a second factor which is now a more general space of the no-scale type, 
with sectional curvature $R=R_{Y,N}$ that is in general non-constant but still quite restricted: 
\be
{\cal M} = \frac {SU(1,1)}{U(1)} \times {\cal M}_{Y,N} \,.
\label{Mgeneral}
\ee
The superpotential is instead, as before, dominated by additional 
unknown effects and can thus a priori take an arbitrary and generic form:
\be
W = W\big(S,T^A\!,\Phi^\alpha\big) \,.
\ee 
This gives a somewhat more realistic picture of the generic situation that might be expected in string models,
which has been less studied so far.

One of the biggest phenomenological concerns in string models is spontaneous supersymmetry breaking. 
A first crucial issue is of course to understand the dynamical origin of this breaking within the microscopic theory. 
We shall however not touch this problem and simply assume very optimistically that while the K\"ahler potential $K$
is approximately fixed the effective superpotential $W$ can be completely generic. This can then lead to a vacuum that 
spontaneously breaks supersymmetry, with arbitrary values for the auxiliary fields of chiral multiplets. For simplicity, 
we shall assume that the values of the auxiliary fields of vector multiplets are negligible, as it naturally tends to be the 
case, and thus ignore vector multiplets altogether. A second important issue is then the structure of the scalar and 
fermion masses. For generic directions in the chiral multiplet space, these two kinds of masses significantly depend 
both on the form of $K$ and $W$, and therefore nothing very predictive can be said from the knowledge of the one 
without the other. It however turns out that while their common supersymmetric part depends more on $W$ than $K$, 
their relative non-supersymmetric splitting depends more on $K$ than $W$. This suggests that one may be able to derive 
some constraints on the structure of these splitting from the knowledge of $K$ without $W$. This is indeed 
the case, as can be seen from the fact that the general expression for this mass splitting depends 
only on the sigma-model Riemann tensor, which characterizes the geometry of the scalar manifold, and the Goldstino 
vector defined by the auxiliary fields, which characterizes the direction of supersymmetry breaking in field space. Assuming 
vanishing and stationary vacuum energy, one finds:
\be
\Delta m^2_{I \bar J} 
= - \Big[R_{I \bar J K \bar L} - \text{\small $\frac 13$} \big(g_{I \bar J} g_{K \bar L} + g_{I \bar L} g_{K \bar J}\big) \Big] 
F^K \bar F^{\bar L} \,.
\ee
For a given $K$ the forms of $g_{I \bar J}$ and $R_{I \bar J K \bar L}$ are fixed, and by varying $W$ one may 
only change the point at which these are evaluated and the vector $F^K$ with which they are contracted.
The vanishing of the cosmological constant moreover implies that $g_{K \bar L} F^K \bar F^{\bar L} = 3\, m_{3/2}^2$,
and for fixed gravitino mass $m_{3/2}$ one may thus vary only the direction defined by the Goldstino unit vector
$f^I = F^I/|F|$. The final observation is now that for those special directions in the chiral multiplet space for which 
supersymmetric mass terms are not allowed and vanish, any information on the non-supersymmetric splitting naturally 
translates into an information on the absolute masses too.

It turns out that there are two particularly important kinds of special directions of the above type in the chiral multiplet space, 
for which not only supersymmetric mass terms are not allowed, but the masses of the corresponding fermions are actually 
fixed while the masses for the corresponding scalars are entirely controlled by the above splitting effect. The first special 
direction is the Goldstino direction $f^I$ within the hidden subsector of fields with non-vanishing VEVs \cite{GRS1,DD}. Indeed, 
in this direction the fermion is the would-be Goldstino mode, which would be forced to have a vanishing mass by the
superGoldstone theorem in the rigid limit and is actually eaten by the gravitino through the superHiggs mechanism 
in the local case. As a result, the average mass of the two real sGoldstino scalar partners is equal to the splitting matrix 
along this direction, and thus reads 
\be
m^2{\!\!\!}_{f \bar f} = \big(2 - 3\, R_{I \bar J K \bar L} \, f^I \hspace{-1pt} \bar f^{\bar J} \!
f^{\hspace{-1pt} K} \hspace{-3pt} \bar f^{\bar L} \big)\, m_{3/2}^2 \,.
\label{mff}
\ee
The second kind of special directions are actually all the flavor directions $v^I$ within the visible subsector of fields with 
vanishing VEVs and standard model quantum numbers \cite{KL,BIM}. Indeed, in this subspace the fermions correspond 
to the ordinary quarks and leptons and are thus not allowed to have direct mass terms, because these are forbidden by the 
standard model gauge symmetries. As a consequence, the mass matrix of the scalars in this sector, which are the squarks and 
sleptons, is equal to the splitting matrix along this subspace, and is therefore given by
\be
m^2_{v \bar v} = \big(1 - 3 \, R_{I \bar J K \bar L} \, v^I \bar v^{\bar J} \! 
f^{\hspace{-1pt} K} \hspace{-3pt} \bar f^{\bar L} \big)\, m_{3/2}^2 \,.
\label{mvv}
\ee
It is evident from the structure of eqs.~(\ref{mff}) and (\ref{mvv}) that the knowledge of the curvature tensor 
$R_{I \bar J K \bar L}$ of the scalar manifold is a quite crucial information. 
Indeed, for $m^2{\!\!}_{f \bar f}$ to be positive one needs the sectional curvature 
$R(f) = - R_{I \bar J K \bar L} f^I \! \bar f^{\bar J} \! f^{\hspace{-1pt} K} \hspace{-3pt} \bar f^{\bar L}$ 
in the plane defined by the vector $f^I$ to be larger than the critical value $-2/3$, while for 
$m^2_{v \bar v}$ to be universal and positive one needs the bisectional curvature 
$R(v,f) = - R_{I \bar J K \bar L} v^I \bar v^{\bar J} \! f^{\hspace{-1pt} K} \hspace{-3pt} \bar f^{\bar L}$ in 
the planes defined by the orthogonal vectors $v^I$ and $f^I$ to be independent of $v^I$ and larger 
than the critical value $-1/3$. For the dilaton sector alone, the situation is fixed 
and one finds a negative contribution to $m^2{\!\!}_{f \bar f}$ and a universal contribution to 
$m^2{\!\!}_{v \bar v}$. For the no-scale sector on its own, there are instead various possibilities. In the simplest case of 
maximally symmetric manifold, one finds a vanishing contribution to $m^2{\!\!}_{f \bar f}$ and a vanishing contribution 
to $m^2{\!\!}_{v \bar v}$ \cite{noscale1,noscale2,noscale3}. In the more general case of non-maximally but still symmetric 
manifolds, it has been shown through an exhaustive study of all the possible cases that the contribution to $m^2{\!\!}_{f \bar f}$ 
is always negative or at best vanishing \cite{GRS2}, while the contribution to $m^2{\!\!}_{v \bar v}$ is always allowed to vanish 
and thus be trivially universal \cite{LPS,AS1}. In the most general case of non-symmetric manifolds one has instead
a richer set of possibilities which were only partly studied so far, both for the contribution to $m^2{\!\!}_{f \bar f}$ \cite{CGGLPS1} 
and the one to $m^2{\!\!}_{v \bar v}$ \cite{AS2}. But one can show that there always exists at least one special 
choice for $f^I$ such that $R(f) = -2/3$ and $R(v,f) = -1/3$ for any $v^I$, leading to vanishing contributions for both
$m^2{\!\!}_{f \bar f}$ and $m^2{\!\!}_{v \bar v}$. 
Notice finally that if one allows the addition of an extra unspecified sector to uplift the cosmological constant along 
the lines of \cite{KKLT}, the situation becomes more flexible and the impact on hidden-sector and visible sector 
scalar masses has been partly studied in \cite{LRSS1,LRSS2} and \cite{CFNO}. But here we will exclude this option 
and only consider those fields that are honestly described by the effective theory.

The aim of this paper is to study the general structure of the curvature for spaces of the general type 
(\ref{Mgeneral}), focusing on the second factor ${\cal M}_{Y,N}$. We will first do this in full generality for 
arbitrary functions $Y$ and $N^A$, and compute both the Riemann tensor and its covariant derivative
in terms of derivatives of these functions, in order to be able to describe how the geometry changes
when departing from the particular case of symmetric manifolds. We will then specialize to the case of 
generic weekly coupled string models based on compactifications on a Calabi-Yau manifold with a 
holomorphic vector bundle over it, where the function $Y$ can be parametrized in terms of the intersection 
numbers $d_{ABC}$ of the Calabi-Yau manifold while the functions $N^A$ can be parametrized in terms of 
some matrices $c^A_{\alpha\beta}$ related to the topology of the gauge bundle over it. We will study both heterotic 
and orientifold constructions, and show that these two classes of models produce scalar manifolds which 
are in some sense dual to each other and coincide whenever they are symmetric manifolds. We will then 
study the conditions that $d_{ABC}$ and $c^A_{\alpha\beta}$ have to satisfy for these manifolds 
to be symmetric, and we will identify two kinds of combinations of these parameters, denoted by $a_{ABCD}$ 
and $b^{AB}_{\alpha\beta}$, which control the departure of such geometries from each other and from the common 
special case of symmetric spaces. We will finally use these results to study the structure of the average 
sGoldstino mass (\ref{mff}) and the soft scalar masses (\ref{mvv}) in such models, as functions of the numbers 
$d_{ABC}$ and $c^A_{\alpha\beta}$, and investigate the conditions under which these may respectively be positive 
and flavor universal, as required by phenomenological considerations. Special attention will be devoted 
to the role of the parameters $a_{ABCD}$ and $b^{AB}_{\alpha\beta}$ that control the deviation from the symmetric 
situation.

The rest of the paper is organized as follows. In section 2 we study the curvature of a generic no-scale manifold. 
In sections 3 and 4 we specialize to the cases of heterotic and orientifold models and work out more 
concretely the form of the curvature tensor and the conditions under which it is covariantly constant. In section 
5 we summarize the general features, similarities and differences of the scalar geometries emerging from these 
two classes of models, and discuss the common case where they are coset spaces. In section 6 we apply 
these results to the study of the average sGoldstino mass and derive the conditions under which this can be 
positive and thus allow for a metastable vacuum. In section 7 we similarly apply the same results to 
the study of the visible soft scalar masses and discuss the conditions under which these can be flavor universal. 
Finally, in section 8 we summarize our results and state our conclusions.

\section{General no-scale manifolds}  
\setcounter{equation}{0}

The scalar manifolds describing the no-scale sector of K\"ahler moduli and matter fields of a string model enjoy, 
as already said, some general features that strongly constrain their geometry. The coordinates $Z^i$ split into two subsets 
$T^A$ and $\Phi^\alpha$, such that the K\"ahler potential takes the following general form:
\be
K = - \log Y(J^A) \,,\;\; J^A = T^A + \bar T^A - N^A(\Phi^\alpha\!, \bar \Phi^\alpha) \,.
\ee
The function $Y$ has the general property of being homogeneous of degree three, that is:
\be
J^A \frac \partial{\partial J^A} Y = 3\, Y\,.
\label{hom}
\ee
The functions $N^A$ may instead be arbitrary, the only crucial property of the variables $J^A$ being that:
\be
\frac \partial{\partial J^A} = \frac \partial{\partial T^A} = \frac \partial{\partial \bar T^A} \,.
\ee

This type of manifolds, called no-scale manifolds, enjoys a series of simple special properties, which have 
far reaching physical consequences. In terms of $K = - \log Y$, the homogeneity constraint (\ref{hom}) implies
that $J^A \partial K/\partial J^A = - 3$.  But $K_i = \partial K/\partial J^A \partial J^A/\partial Z^i$,
and in particular $K_A = \partial K/\partial J^A$, so that this homogeneity constraint can be rewritten 
in the following simple form:
\be
K_A J^A = - 3 \,.
\ee 
Taking a derivative of this relation, it follows that $K_{i A} J^A + K_{i} = 0$. 
Acting then on this with the inverse metric one deduces that 
\be
K^i = - \delta^i_A J^A \,.
\ee
Form these two properties it finally follows that there is a no-scale structure, namely:
\be
K_i K^i = 3 \,.
\label{noscalegen}
\ee

\subsection{Geometry}

Let us now examine the impact of these properties on the geometry. To start with, it is 
convenient to write down what happens for a generic K\"ahler manifold where the K\"ahler potential 
is parametrized as $K = - \log Y$, with completely arbitrary $Y$. It is straightforward to show that 
$K_i = - Y_i/Y$ and $K^i = (1 - Y_p Y^{\inv p \bar q}Y_{\bar q}/Y)^{-1} Y^{\inv i \bar \jmath} Y_{\bar \jmath}$,
and that the metric and its inverse can be written as
\bea
g_{i \bar \jmath} \b=\b - \frac{Y_{i \bar \jmath}}{Y} + \frac{Y_i Y_{\bar \jmath}}{Y^2} \,, \label{g} \\
g^{i \bar \jmath} \b=\b - Y Y^{\inv i \bar \jmath} - \bigg(1 - \frac {Y_p Y^{\inv p \bar q}Y_{\bar q}}Y \bigg)^{\!\!-1}\, 
Y^{\inv i \bar r} Y_{\bar r} Y^{\inv \bar \jmath s} Y_s  \,. \label{ginv}
\eea
The Christoffel symbols are instead given by:
\be
\Gamma_{ij \bar k} = - \frac{Y_{i j \bar k}}{Y} + \frac{ Y_{ij} Y_{\bar k}}{Y^2}
- \frac {g_{i \bar k} Y_j + g_{j \bar k} Y_i}Y \,.
\label{Chris}
\ee
The Riemann tensor is then found to read:
\bea
R_{i \bar \jmath p \bar q} \b=\b g_{i \bar \jmath} \, g_{p \bar q} + g_{i \bar q} \, g_{p \bar \jmath} 
- \frac {Y_{i \bar \jmath p \bar q}}{Y} - \frac {Y_{i p \bar s} {Y^{\bar s}}_{\bar \jmath \bar q}}{Y^2} \nn \\
\b\;\b +\, \bigg(1 - \frac {Y_p Y^{\inv p \bar q}Y_{\bar q}}Y \bigg)^{\!\!-1} \frac {Y_{i p} Y_{\bar \jmath \bar q}}{Y^2} 
+ \frac {Y_{i p} Y_{\bar \jmath \bar q r} Y^r}{Y^3}
+ \frac {Y_{\bar \jmath \bar q} Y_{i p \bar s} Y^{\bar s}}{Y^3} \,.
\label{Riem}
\eea
Finally, one may try to compute the covariant derivative of the Riemann tensor, but this produces a 
quite complicated expression:
\bea
\nabla_{\!n} R_{i \bar \jmath p \bar q} \b=\b \text{complicated expression} \,.
\eea

Let us now see what happens if the manifold is of the no-scale type discussed above. 
The generic no-scale property $K_i K^i = 3$ has a very simple and straightforward consequence. 
Indeed, it implies that $(Y_i Y^{\inv i \bar \jmath} Y_{\bar \jmath}/Y)(1 - Y_p Y^{\inv p \bar q}Y_{\bar q}/Y)^{-1} = -3$,
that is:
\be
Y_i Y^{i \bar \jmath} Y_{\bar \jmath} = \text{\small $\frac 32$}\, Y \,.
\label{impnoscale}
\ee
The more specific homogeneity properties $K_A J^A = - 3$ and $K^i = - \delta^i_A J^A$, which also 
imply the no-scale property $K_i K^i = 3$, have further consequences. 
They respectively imply that $Y_A J^A = 3\, Y$ and $Y^{i \bar r} Y_{\bar r} = 1/2\,\delta^i_A J^A$, which 
also give the no-scale property $Y_i Y^{i \bar \jmath} Y_{\bar \jmath} = 3/2\, Y$. To work out the most 
relevant implications of these relations, we may start from the no-scale relation written as $Y_l K^l = - 3\, Y$, 
and deduce by taking further derivatives and making use of the homogeneity properties that 
$Y_{i l} K^l = - 2\, Y_i$, $Y_{i j l} K^l = - Y_{i j}$, $Y_{i j k l } K^l = 0$ and
$Y_{\bar \imath l} K^l = - 2\, Y_{\bar \imath}$, $Y_{\bar \imath \bar \jmath l} K^l = - Y_{\bar \imath \bar \jmath}$,
$Y_{\bar \imath \bar \jmath \bar k l } K^l = 0$. Finally, recalling that $K_i = - Y_i/Y$ and thus $K^i = - Y^i/Y$, 
one arrives at the following relations:
\bea
\a\a Y_l Y^l = 3\, Y^2 \,, \\
\a\a Y_{i l} Y^l = 2\, Y Y_i \,,\;\; Y_{\bar \imath l} Y^l = 2\, Y Y_{\bar \imath} \,, \\
\a\a Y_{i j l} Y^l = Y Y_{i j} \,,\;\; Y_{\bar \imath \bar \jmath l} Y^l = Y Y_{\bar \imath \bar \jmath} \,, \label{imphomogen} \\
\a\a Y_{i j k l} Y^l = 0 \,,\;\; Y_{\bar \imath \bar \jmath \bar k l} Y^l = 0 \,.
\eea
By taking various derivatives of these relations, one can also further deduce that:
\bea
\a\a Y_{i \bar \jmath l} Y^l = - 2\, g_{i \bar \jmath} + Y_{i k} Y^k{}_{\bar \jmath} \,, \\
\a\a Y_{i j \bar k l} Y^l = Y Y_{i j \bar k} - Y_{i j} Y_{\bar k} + Y_{i j l} Y^l{}_{\bar k} \,,\\[-0.5mm]
\a\a Y_{i j}{}^l Y_{l k} = - Y Y_{i j k} + Y_{ij} Y_k \,, \\[0.2mm]
\a\a Y Y_{i \bar \jmath \bar k l} Y^l = Y^2 Y_{i \bar \jmath \bar k} + Y Y_i Y_{\bar \jmath \bar k} 
- Y_{\bar \jmath \bar k l} Y^{l m} Y_{m p} \,.
\eea

Coming back to the geometry, we now see that the relation (\ref{impnoscale}) descending from the 
no-scale property implies that the factor $(1 - Y_p Y^{\inv p \bar q}Y_{\bar q}/Y)^{-1}$ appearing in 
(\ref{ginv}) and (\ref{Riem}) simplifies to a constant and is equal to $-2$, while the 
properties (\ref{imphomogen}) descending from homogeneity further imply that the last three 
terms of (\ref{Riem}) cancel out. To sum up, one then has $K_i = - Y_i/Y$ and 
$K^i = - 2\, Y^{\inv i \bar \jmath} Y_{\bar \jmath} = - \delta^i_A J^A$, and the metric and its inverse read:
\bea
g_{i \bar \jmath} \b=\b - \frac{Y_{i \bar \jmath}}{Y} + \frac{Y_i Y_{\bar \jmath}}{Y^2} \,, \\
g^{i \bar \jmath} \b=\b - Y Y^{i \bar \jmath} + 2\, Y^{\inv i \bar r} Y_{\bar r} Y^{\inv \bar \jmath s} Y_s \,.
\eea
The Christoffel symbols are as before
\be
\Gamma_{ij \bar k} = - \frac{Y_{i j \bar k}}{Y} + \frac{ Y_{ij} Y_{\bar k}}{Y^2}
- \frac {g_{i \bar k} Y_j + g_{j \bar k} Y_i}Y \,.
\ee
The Riemann tensor is instead given by the following very simple expression:
\bea
R_{i \bar \jmath p \bar q} \b=\b g_{i \bar \jmath} \, g_{p \bar q} + g_{i \bar q} \, g_{p \bar \jmath} 
- \frac {Y_{i \bar \jmath p \bar q}}{Y} - \frac {Y_{i p \bar s} {Y^{\bar s}}_{\bar \jmath \bar q}}{Y^2} \,.
\label{Riemnoscale}
\eea
Finally, after a straightforward computation using several of the above identities, the 
covariant derivative of the curvature tensor is found to be completely symmetric in its 
holomorphic and antiholomorphic indices as implied by the Bianchi identity and reads
\bea
\nabla_{\!n} R_{i \bar \jmath p \bar q} \b=\b 
g_{n \bar \jmath} \Gamma_{i p \bar q} + g_{i \bar \jmath} \Gamma_{n p \bar q} + g_{p \bar \jmath} \Gamma_{i n \bar q} 
+ g_{n \bar q} \Gamma_{i p \bar \jmath} + g_{i \bar q} \Gamma_{n p \bar \jmath} + g_{p \bar q} \Gamma_{i n \bar \jmath} 
\nn \\[2mm]
\b\;\b -\, \Gamma_{n i}{}^k R_{k \bar \jmath p \bar q} - \Gamma_{n p}{}^k R_{k \bar \jmath i \bar q}
- \Gamma_{n p}{}^k R_{k \bar \jmath i \bar q} 
- \frac {Y_n R_{i \bar \jmath p \bar q} + Y_i R_{n \bar \jmath p \bar q} + Y_p R_{i \bar \jmath n \bar q}}{Y} \nn \\[1mm]
\b\;\b +\, \frac {Y_n g_{i \bar \jmath} g_{p \bar q} + Y_i g_{n \bar \jmath} g_{p \bar q} + Y_p g_{i \bar \jmath} g_{n \bar q}
+ Y_n g_{i \bar q} g_{p \bar \jmath} + Y_i g_{n \bar q} g_{p \bar \jmath} + Y_p g_{i \bar q} g_{n \bar \jmath}}{Y}  \nn \\
\b\;\b -\, \frac {Y_{n i p \bar q \bar \jmath}}{Y} 
- \frac {Y_{nip} Y_{\bar q \bar \jmath}}{Y^2} - \frac {Y_{nip}{}^k Y_{k\bar q \bar \jmath}}{Y^2} \,.
\eea
The above expressions represent novel general results for the form of the geometry of no-scale manifolds, which are 
rather similar in form to those holding for special K\"ahler manifolds. We see that generic no-scale manifolds 
are neither homogeneous nor symmetric, since a priori $\nabla_n R_{i \bar \jmath p \bar q} \neq 0$.
On the other hand, there exist particular no-scale manifolds which are coset spaces, and correspondingly 
$\nabla_n R_{i \bar \jmath p \bar q} = 0$, but clearly there can be only a finite set of these, and this must be 
a subset of all the possible K\"ahler coset manifolds described in \cite{CV}.

\subsection{General properties}

Using the various identities listed above, it is straightforward to show that all the geometrical quantities 
that have been computed enjoy some universal properties along the special direction defined by the unit 
vector
\be
k^i = - \text{\small $\frac 1{\sqrt{3}}$}\, K^i \,.
\ee
For the metric on has
\bea
\a\a g_{i \bar \jmath}\, \bar k^{\bar \jmath} = \bar k_i \,, \label{genbegin} \\
\a\a g_{i \bar \jmath}\, k^i \bar k^{\bar \jmath} = 1 \,,
\eea
for the Christoffel symbols
\bea
\a\a \Gamma_{i j \bar k}\, \bar k^{\bar k} = - \text{\small $\frac 2{\sqrt{3}}$} \, K_{ij} \,,\;\; 
\Gamma_{i j \bar k}\, k^j = - \text{\small $\frac 1{\sqrt{3}}$} \big(g_{i \bar k} + K_{i l} K^l{}_{\bar k}\big) \,, \\[-1mm]
\a\a \Gamma_{i j \bar k}\, k^j \bar k^{\bar k} = - \text{\small $\frac 2{\sqrt{3}}$}\, \bar k_i \,, \\[-1mm]
\a\a \Gamma_{i j \bar k}\, k^i k^j \bar k^{\bar k} = - \text{\small $\frac 2{\sqrt{3}}$} \,,
\eea
for the Riemann tensor
\bea
\a\a R_{i \bar \jmath p \bar q}\, \bar k^{\bar q} = - \text{\small $\frac 1{\sqrt{3}}$} \Gamma_{i p \bar \jmath} \,, \\[-1mm]
\a\a R_{i \bar \jmath p \bar q}\, k^p \bar k^{\bar q} = \text{\small $\frac 13$} \big(g_{i \bar \jmath} 
+ K_{i l} K^l{}_{\bar \jmath}\big) \,,\;\;
R_{i \bar \jmath p \bar q}\, \bar k^{\bar \jmath} \hspace{0.5pt} \bar k^{\bar q} = \text{\small $\frac 23$}\, K_{i p} \,, \\
\a\a R_{i \bar \jmath p \bar q}\, \bar k^{\bar \jmath}   \hspace{0.5pt} k^p \bar k^{\bar q} = \text{\small $\frac 23$}\, \bar k_i \,, \\
\a\a R_{i \bar \jmath p \bar q}\, k^i \bar k^{\bar \jmath} \hspace{0.5pt} k^p \bar k^{\bar q} = \text{\small $\frac 23$} \,, \label{genend}
\eea
and finally for the covariant derivative of the Riemann tensor
\bea
\a\a \nabla_{\!n} R_{i \bar \jmath p \bar q} \, k^p = 0 \,,\;\; 
\nabla_{\!n} R_{i \bar \jmath p \bar q} \, \bar k^{\bar q} = - \text{\small $\frac 1{\sqrt{3}}$} \big(
\nabla_{\!n} \Gamma_{i \bar \jmath p} + R_{i \bar \jmath p \bar q} K^{\bar q}{}_s\big) \,, \\[-1mm]
\a\a \nabla_{\!n} R_{i \bar \jmath p \bar q} \, k^i k^p = 0  \,,\;\; 
\nabla_{\!n} R_{i \bar \jmath p \bar q} \, k^p \bar k^{\bar q} = 0 \,,\;\; 
\nabla_{\!n} R_{i \bar \jmath p \bar q}\, \bar k^{\bar \jmath} \hspace{0.5pt} \bar k^{\bar q} = 0 \,, \\[1mm]
\a\a \qquad \cdots \nn \\
\a\a \nabla_{\!n} R_{i \bar \jmath p \bar q}\, k^n \hspace{-0.5pt} k^i \bar k^{\bar \jmath} \hspace{0.5pt} k^p \bar k^{\bar q} = 0 \,.
\eea

With the help of the above properties, it is now straightforward to demonstrate that the sectional curvature 
in the plane defined by the special vector $k^i$ is equal to $-2/3$ and that the bisectional curvature 
in the planes defined by the special vector $k^i$ and any orthogonal vector $v^i$ in the visible subsector 
is equal to $-1/3$, as anticipated in the introduction: $R(k) = -2/3$ and $R(v,k) = -1/3$.

\section{Heterotic models}
\setcounter{equation}{0}

Let us now consider the more specific case of heterotic string models compactified on a Calabi-Yau 
manifold $X$ with a holomorphic vector bundle over it \cite{CYhet1,CYhet2,CYhet3}. In this setting, the 
K\"ahler moduli $T^A$ are associated to harmonic $(1,1)$ forms $\omega_A$ on $X$, which are dual 
to the harmonic $(2,2)$ forms $\omega^A$ on $X$, while the matter fields $\Phi^\alpha$ are associated to 
bundle-valued harmonic $(1,0)$ forms $u_\alpha$ on $X$. The relevant numbers defining the low-energy 
effective theory are then given by the following integrals:
\bea
\a\a d_{ABC} = \int_X \omega_A \wedge \omega_B \wedge \omega_C \,, \\
\a\a c^A_{\alpha\beta} = \int_X \omega^A \wedge {\rm tr}\, [u_\alpha \wedge \bar u_\beta] \,.
\eea
We will take the point of view that a priori $d_{ABC}$ can be an arbitrary symmetric symbol
and similarly that $c^A_{\alpha\beta}$ can be an arbitrary set of Hermitian matrices, and study the low-energy
effective scalar geometry as a function of these parameters.

\subsection{K\"ahler potential}

The effective K\"ahler potential for the K\"ahler moduli $T^A$ and the matter fields $\Phi^\alpha$ can be worked 
out by dimensionally reducing the kinetic terms of the ten-dimensional supergravity theory describing 
the heterotic string below the Planck scale down to four dimensions, retaining only the harmonic components 
of the fields. The full moduli dependence was worked out in \cite{CFG,CdlO}, and the leading matter field 
dependence in \cite{DKL}, while the full matter field dependence was studied only more recently in 
\cite{AS2,PS} (see also \cite{GLM,BLM,BO}), generalizing the results that were available from \cite{FKP} for the 
special case of orbifold limits. The complete result depends on the parameters $d_{ABC}$ and $c^A_{\alpha\beta}$ 
and takes the form $K = - \log Y$, where the function $Y$ depends only on certain combinations of fields. 
More precisely, we have 
\be
Y = {\cal V} \,,
\ee
where ${\cal V}$ denotes the volume of the Calabi-Yau manifold and is given by the following 
expression in terms of the real geometric moduli fields $v^A$:
\be
{\cal V} = \text{\small $\frac 16$}\, d_{ABC} v^A v^B v^C \,.
\label{Vhet}
\ee
The real fields $v^A$ are then linked to the following real combination of complex fields:
\be
J^A = T^A + \bar T^A - c^A_{\alpha \beta} \Phi^\alpha \bar \Phi^\beta \,.
\ee
The relation between the $v^A$ and the $J^A$ is in this case trivial and given simply by
\be
J^A = v^A \,.
\label{Jvhet}
\ee
It follows that $Y$ can be explicitly written in terms of the variables $J^A$ and simply reads
\be
Y = \text{\small $\frac 16$}\, d_{ABC} J^A J^B J^C \,.
\label{Yhet}
\ee
We see that $Y$ is a homogenous function of degree three in the variables $J^A$, and we therefore 
have a no-scale manifold. Moreover, the function $Y$ is in this case a simple cubic polynomial 
in the variables $J^A$, involving only integer powers of them.

\subsection{Canonical parametrization}

We now want to study the above space at a given reference point, which can be thought of as the one 
defined by the VEVs $\langle T^A \rangle$ and $\langle \Phi^\alpha \rangle$ that the scalar fields eventually acquire 
in the presence of a non-trivial superpotential. For simplicity, we shall restrict to the situation where the moduli 
have sizable VEVs whereas the matter fields have negligible VEVs, that is:
\be 
\langle T^A \rangle \neq 0 \,,\;\; \langle \Phi^\alpha \rangle = 0 \,.
\label{refhet}
\ee

For any given reference point of the type (\ref{refhet}), it is possible to define a particularly convenient 
canonical parametrization, in such a way as to simplify the form of the metric, the Christoffel symbols and 
the curvature tensor at that point. To this aim, we proceed along the lines of 
\cite{GST,CKVDFdWG} and consider the field redefinitions
\be
\hat T^A = U^A_{\;\;B} T^B \,,\;\; \hat \Phi^\alpha = V^\alpha_{\;\;\beta} \Phi^\beta \,,
\ee
together with the parameter redefinitions
\be
\hat d_{ABC} = \alpha \, U_{\;\;\;\;\;A}^{\text{-1}D} U_{\;\;\;\;\;B}^{\text{-1}E} U_{\;\;\;\;\;C}^{\text{-1}F} d_{DEF} \,,\;\;
\hat c^A_{\alpha\beta} = U^A_{\;\;B} V_{\;\;\;\;\;\;\alpha}^{\text{-1}\gamma} \bar V_{\;\;\;\;\;\beta}^{\text{-1}\delta} 
c^B_{\gamma\delta} \,.
\ee
Under the above combined transformations, with $U^A_{\;\;B}$ a real matrix, $V^\alpha_{\;\;\beta}$ a complex matrix 
and $\alpha$ a positive real number, the real geometrical moduli transform simply as $\hat v^A = U^A_{\;\;B} v^B$ 
and the K\"ahler potential remains unchanged, modulo an irrelevant K\"ahler transformation:
\be
\hat K = K - \log \alpha \,.
\ee 
We may now choose $U^A_{\;\;B}$ and $V^\alpha_{\;\;\beta}$ in such a way that the VEVs of the fields are aligned along 
just one direction, the VEV of the metric becomes diagonal, and the overall scale of these two quantities is set to 
some reference value. We may furthermore choose $\alpha$ to set the overall scale of the intersection numbers 
to a convenient value. More specifically, we shall require that in the new basis the reference point should be at
\be
\langle \hat T^A \rangle = \text{\small $\frac {\sqrt{3}}2$} \, \delta^A_0 + i (\cdots)\,,\;\; \langle \hat \Phi^\alpha \rangle = 0 \,,
\ee
the metric at that point should take the form
\be
\langle \hat g_{A \bar B} \rangle = \delta_{AB} \,,\;\;
\langle \hat g_{\alpha \bar \beta} \rangle = \delta_{\alpha\beta}\,,\;\;
\langle \hat g_{A \bar \beta} \rangle = 0 \,,
\ee
and finally the K\"ahler frame should be such that at that point
\be
\langle \hat K \rangle = 0 \,.
\ee
It is easy to get convinced by a counting of parameters that it is indeed always possible to 
impose this kind of conditions. Moreover, by comparing the expressions for the 
VEVs of the fields, the metric and the K\"ahler potential with the values required in the previous 
equations, we deduce that the new values of the numerical coefficients $\hat d_{ABC}$ and 
$\hat c^A_{\alpha\beta}$ must take the following form:
\bea
\a\a \hat d_{000} = \text{\small $\frac 2{\sqrt{3}}$} \,,\;\; \hat d_{00a} = 0 \,,\;\; 
\hat d_{0 ab} = - \text{\small $\frac 1{\sqrt{3}}$}\, \delta_{ab} \,,\;\; \hat d_{abc} = \text{generic} \,, \label{dcanor} \\
\a\a \hat c^0_{\alpha\beta} = \text{\small $\frac 1{\sqrt{3}}$} \, \delta_{\alpha\beta} \,,\;\; 
\hat c^a_{\alpha\beta} = \text{generic} \label{lambdacanor} \,.
\eea
Notice finally that from the point of view of the Calabi-Yau manifold, the canonical frame just corresponds 
to a convenient choice of basis for harmonic forms, which is suitably oriented with respect to the K\"ahler 
form and normalized in such a way as to get unit volume, since $\langle \hat v^A \rangle = \sqrt{3}\, \delta^A_{0}$
and $\langle \hat {\cal V} \rangle = 1$. Moreover, by comparing with the general results of section 2 we see that 
the canonical frame essentially corresponds to choosing a parametrization such that the special direction $k^i$ 
is identified with one of the fields, since $\langle \hat k^i \rangle = \delta^i_0$. 

\subsection{Geometry}

Let us now explore the geometry at a given reference point by using the new canonical coordinates.
For notational simplicity, we drop from now on the hats referring  to the definition of this special frame, and
also the brackets referring to the special point. 

We start by computing the first five partial derivatives of ${\cal V}$, which are the basic ingredients that 
we need. It is convenient to introduce the following notation:
\bea
\a\a d_{AB} = d_{ABC} J^C \,, \\
\a\a d_A = \text{\small $\frac 12$} \, d_{ABC} J^B J^C \,.
\eea
In terms of these quantities, one easily finds
\bea
\a\a {\cal V}_i = d_{A} J^A_i \,, \\[1mm]
\a\a {\cal V}_{i \bar \jmath} =  d_{AB} J^A_i J^B_{\bar \jmath} +  d_A J^A_{i \bar \jmath} \,, \\[1mm]
\a\a {\cal V}_{i \bar \jmath p} =  d_{ABC} J^A_i J^B_{\bar \jmath} J^C_p
+ d_{AB} \big(J^A_{i \bar \jmath} J^B_p + J^A_{p \bar \jmath} J^B_{i} \big) \,, \\[1mm]
\a\a {\cal V}_{i \bar \jmath p \bar q} =  d_{ABC} \big(J^A_{i \bar \jmath} J^B_p J^C_{\bar q} + \mbox{3 p.} \big)
+ d_{AB} \big(J^A_{i \bar \jmath} J^B_{p \bar q} \!+\! J^A_{i \bar q} J^B_{p \bar \jmath} \big) \,, \\[1.2mm]
\a\a {\cal V}_{i \bar \jmath p \bar q n} = d_{ABC} \big(J^A_{i \bar \jmath} J^B_{p \bar q} J^C_n 
+ \mbox{5 p.}\big) \,.
\eea
Using these expressions and going to the canonical frame at the reference point, one then obtains
the following non-vanishing entries for the derivatives of $Y={\cal V}$:
\bea
\a\a Y_A = d_A \,, \\[0.5mm]
\a\a Y_{A \bar B} = d_{AB} \,,\;\; Y_{\alpha \bar \beta} = - \delta_{\alpha\beta} \,, \\[0.5mm]
\a\a Y_{A \bar B C} = d_{AB C} \,,\;\; Y_{A \alpha \bar \beta} = - d_{AX} c^X_{\alpha\beta} \,, \\[0.5mm]
\a\a Y_{A \bar B C \bar D} = 0 \,,\;\; Y_{A \bar B \alpha \bar \beta} = - d_{ABX} c^X_{\alpha\beta} \,,\;\;
Y_{\alpha \bar \beta \gamma \bar \delta} = d_{XY} \big(c^X_{\alpha\beta} c^Y_{\gamma\delta} 
+ c^X_{\alpha\delta} c^Y_{\gamma\beta} \big) \,, \\[0.5mm]
\a\a Y_{A \bar B C \bar D E} = 0 \,,\;\; Y_{A \bar B C \alpha \bar \beta} = 0 \,,\;\;
Y_{A \alpha \bar \beta \gamma \bar \delta} = d_{AXY} \big(c^X_{\alpha\beta} c^Y_{\gamma\delta} 
+ c^X_{\alpha\delta} c^Y_{\gamma\beta} \big) \,.
\eea
The parameters in these expressions are now given by 
$d_A = 3/2\, d_{A00}$, $d_{AB} = \sqrt{3}\, d_{AB0}$ and $d_{ABC}$ taking the already 
studied restricted form. Moreover, it will also be useful to define the following combinations 
of the parameters $d_{ABC}$ and $c^A_{\alpha\beta}$:
\bea
\a\a a_{ABCD} = \text{\small $\frac 12$} \big(d_{ABX} d_{XCD} \!+ d_{ADX} d_{XBC} \!+ d_{ACX} d_{XBD} \big) \nn \\
\a\a \hspace{48pt} -\, \text{\small $\frac 12$} \big(d_{AB} d_{CD} \!+ d_{AD} d_{BC} \!+ d_{AC} d_{BD} \big) \nn \\
\a\a \hspace{48pt} +\, \text{\small $\frac 12$} \big(d_A d_{BCD} + d_B d_{ACD} + d_C d_{ABD} + d_D d_{ABC} \big) \,, 
\label{ahet} \\
\a\a b^{AB}_{\alpha\beta} = \text{\small $\frac 12$} \Big(\big\{c^A,c^B\big\}_{\hspace{-1pt}\alpha\beta} 
- d_{ABX} c^X_{\alpha\beta} + d_{AB} \delta_{\alpha\beta} - d_A c^B_{\alpha\beta} - d_B c^A_{\alpha\beta} \Big) \,, 
\label{bhet}\\
\a\a \tau^A_{\alpha\beta\gamma\delta} = \text{\small $\frac 12$} \Big(\big[c^A,c^X\big]_{\hspace{-1pt}\alpha\beta} c^X_{\gamma\delta} 
\!+\! \big[c^A,c^X\big]_{\hspace{-1pt}\alpha \delta} c^X_{\gamma \beta} 
\!+\! \big[c^A,c^X\big]_{\hspace{-1pt}\gamma\delta} c^X_{\alpha\beta} 
\!+\! \big[c^A,c^X\big]_{\hspace{-1pt}\gamma\beta} c^X_{\alpha\delta} \Big) \,. 
\label{tauhet}
\eea
These quantities are completely symmetric in their indices of type $A,B,\cdots$ and vanish if one of these
is equal to $0$, meaning that $a_{ABCX} d_X = 0$, $b^{AX}_{\alpha\beta} d_X = 0$ and 
$\tau^X_{\alpha\beta\gamma\delta} d_X = 0$.

Starting from the above expressions, it is now straightforward to compute all the geometric quantities we are interested in.
The metric is trivial and its non-vanishing entries are
\bea
\a\a g_{A \bar B} = \delta_{AB} \,,\\
\a\a g_{\alpha \bar \beta} = \delta_{\alpha\beta} \,.
\eea
The Christoffel connection is instead non-trivial, and its non-vanishing entries are
\bea
\a\a \Gamma_{A B \bar C} = - d_{ABC} + \big(d_{AB} d_C + \mbox{2 p.} \big) - 2\, d_A d_B d_C \,, \\
\a\a \Gamma_{A \alpha \bar \beta} = - c^A_{\alpha\beta} \,.
\eea
The Riemann curvature tensor is found to be given by
\bea
\a\a R_{A \bar B C \bar D} = \delta_{AB} \delta_{CD} + \delta_{AD} \delta_{BC} - d_{ACX} d_{XBD} \,, \\[0.5mm]
\a\a R_{\alpha \bar \beta \gamma \bar \delta} = c^X_{\alpha\beta} c^X_{\gamma\delta} + c^X_{\alpha\delta} c^X_{\gamma\beta} \,, \\
\a\a R_{\hspace{-1pt} \alpha \bar \beta A \bar B} = \delta_{AB} \delta_{\alpha\beta} 
- d_{AX} d_{BY} (c^X c^Y)_{\alpha\beta} + d_{ABX} c^X_{\alpha\beta} \,.
\eea
Finally, the covariant derivative of the Riemann tensor reads
\bea
\a\a \nabla_{\!A} R_{B \bar C D \bar E} = - 2\, a_{ABDX} d_{X C E} \,, \\[1mm]
\a\a \nabla_{\!A} R_{B \bar C \alpha \bar \beta} = a_{ABCX} c^X_{\alpha\beta} 
+ 2\, d_C b^{AB}_{\alpha\beta} - d_{ABX} b^{XC}_{\alpha\beta} + d_{ACX} b^{XB}_{\alpha\beta} 
+ d_{BCX} b^{XA}_{\alpha\beta}  \nn \\[1mm]
\a\a \hspace{62pt} - \big\{b^{AB}, c^C\big\}_{\hspace{-1pt} \alpha \beta} 
+ \big[b^{AC}, c^B\big]_{\hspace{-1pt} \alpha\beta} + \big[b^{BC}, c^A\big]_{\hspace{-1pt} \alpha\beta} \,, \\[0.5mm]
\a\a \nabla_{\!A} R_{\alpha \bar \beta \gamma \bar \delta} = \tau^A_{\alpha\beta\gamma\delta} 
+ b^{AX}_{\alpha\beta} c^X_{\gamma\delta} + b^{AX}_{\alpha\delta} c^X_{\gamma\beta} 
+ b^{AX}_{\gamma\delta} c^X_{\alpha\beta} + b^{AX}_{\gamma\beta} c^X_{\alpha\delta}  \,.
\eea
We thus see that the manifold is a symmetric space with covariantly constant curvature if and only if the quantities 
$a_{ABCD}$, $b^{AB}_{\alpha\beta}$ and $\tau^A_{\alpha\beta\gamma\delta}$ identically vanish.

\section{Orientifold models}
\setcounter{equation}{0}

Let us now consider the other specific case of orientifold string models based on a Calabi-Yau
manifold  $X$ with D7-branes supporting a non-trivial vector bundle and wrapping on some four-cycles 
$C$ of $X$ (see \cite{CYor1,CYor2} for a review). In this case, the K\"ahler moduli $T^A$ are associated to harmonic 
$(1,1)$ forms $\omega^A$ on $X$ which are dual to the harmonic $(2,2)$ forms $\omega_A$, while 
the matter fields $\Phi^\alpha$ are associated to bundle-valued harmonic $(1,0)$ forms $u_\alpha$ 
on $C \subset X$. Notice that for later convenience we use here opposite conventions compared to the 
heterotic case for the position of the index labeling harmonic forms and their duals. Denoting by $i$ the 
embedding map defining $C$ in $X$ and by $i^*$ its pullback on forms, the relevant numbers defining 
the low-energy effective theory are then given by the following integrals: 
\bea
\a\a d^{ABC} = \int_X \omega^A \wedge \omega^B \wedge \omega^C \,, \\
\a\a c^A_{\alpha\beta} = \int_C i^* \omega^A \wedge {\rm tr}\, [u_\alpha \wedge \bar u_\beta] \,.
\eea
We will again take the point of view that a priori $d^{ABC}$ can be an arbitrary symmetric symbol
and similarly that $c^A_{\alpha\beta}$ can be an arbitrary set of Hermitian matrices, and study the low-energy
effective scalar geometry as a function of these parameters.

\subsection{K\"ahler potential}

The effective K\"ahler potential for the K\"ahler moduli $T^A$ and the matter fields $\Phi^\alpha$ can, as before, 
be worked out by dimensionally reducing the kinetic terms of the ten-dimensional supergravity theory 
describing the unoriented string below the Planck scale down to four dimensions, retaining only 
the harmonic components of the fields. The full moduli and matter field dependence was worked out in 
\cite{Korient1,Korient2,Korient3}, generalizing the results that were previously known for the special case of 
orbifold limits (see for example \cite{CYor2}). The complete result depends on the parameters 
$d^{ABC}$ and $c^A_{\alpha\beta}$ and takes again the form $K = - \log Y$, where the function $Y$ depends 
only on certain combinations of fields. More precisely, we have in this case
\be
Y = {\cal V}^2 \,,
\ee
where ${\cal V}$ denotes the volume of the Calabi-Yau manifold and is given by the following 
expression in terms of the real geometric moduli fields $v_A$:
\be
{\cal V} = \text{\small $\frac 16$}\, d^{ABC} v_A v_B v_C \,.
\label{Vor}
\ee
The real fields $v_A$ are then linked to the following real combination of complex fields:
\be
J^A = T^A + \bar T^A - c^A_{\alpha\beta} \Phi^\alpha \bar \Phi^\beta \,.
\ee
The relation between the $v_A$ and the $J^A$ is in this case non-trivial and defined by
the following equation:
\be
J^A = \frac {\partial {\cal V}}{\partial v_A} = \text{\small $\frac 12$}\, d^{ABC} v_B v_C \,.
\label{Jvor}
\ee
It follows that in general $Y$ cannot be explicitly written in terms of the variables $J^A$ 
and is only implicitly defined:
\be
Y = Y(J) \,.
\label{Yor}
\ee
We see that $Y$ is as before a homogenous function of degree three in the variables $J^A$, and we 
therefore have again a no-scale manifold. However, the function $Y$ is in this case no-longer 
always a simple cubic polynomial in the variables $J^A$, and generically involves non-integer powers 
of them.

\subsection{Canonical parametrization}

We now want to study the above space at a given reference point, corresponding to the VEVs 
$\langle T^A \rangle$ and $\langle \Phi^\alpha \rangle$ that the scalar fields eventually acquire. 
For simplicity, we shall again restrict to the situation where the moduli have sizable VEVs 
whereas the matter fields have negligible VEVs, that is:
\be 
\langle T^A \rangle \neq 0 \,,\;\; \langle \Phi^\alpha \rangle = 0 \,.
\label{refor}
\ee

For any given reference point of the type (\ref{refor}), it is again possible to define a particularly convenient 
canonical parametrization, in such a way as to simplify geometrical quantities at that point. 
To this aim, we proceed along the same lines as before and consider the 
field redefinitions
\be
\hat T^A = U^A{\!}_{B} T^B \,,\;\; \hat \Phi^\alpha = V^\alpha{\!}_\beta \Phi^\beta \,,
\ee
together with the parameter redefinitions
\be
\hat d^{ABC} = \alpha^{\inv} U^A{\!}_D U^B{\!}_E U^C{\!}_F d^{DEF} \,,\;\;
\hat c^A_{\alpha\beta} = U^A{\!}_B V^{\inv \gamma}{\!}_\alpha \bar V^{\inv \delta}{\!}_\beta c^B_{\gamma\delta} \,.
\ee
Under the above combined transformations, with $U^A_{\;\;B}$ a real matrix, $V^\alpha_{\;\;\beta}$ a complex 
matrix, and $\alpha$ a positive real number, the real geometrical moduli transform as 
$\hat v_A = \sqrt{\alpha}\, U^{\inv B}{\!}_A v_B$, and the K\"ahler potential 
remains unchanged, modulo an irrelevant K\"ahler transformation:
\be
\hat K = K - \log \alpha \,.
\ee 
We may now choose $U^A_{\;\;B}$ and $V^\alpha_{\;\;\beta}$ such that the VEVs of the fields are aligned along 
just one direction, the VEV of the metric becomes diagonal, and the overall scale of these two quantities is set to 
some reference value. We may furthermore choose $\alpha$ to set the overall scale of the intersection numbers 
to a convenient value. More specifically, we shall require as before that in the new basis the reference point should 
be at
\be
\langle \hat T^A \rangle = \text{\small $\frac {\sqrt{3}}2$} \, \delta^A_0 + i (\cdots)\,,\;\; \langle \hat \Phi^\alpha \rangle = 0 \,,
\ee
the metric at that point should take the form
\be
\langle \hat g_{A \bar B} \rangle = \delta_{AB} \,,\;\;
\langle \hat g_{\alpha \bar \beta} \rangle = \delta_{\alpha\beta}\,,\;\;
\langle \hat g_{A \bar \beta} \rangle = 0 \,,
\ee
and finally the K\"ahler frame should be such that at that point
\be
\langle \hat K \rangle = 0 \,.
\ee
It is again easy to get convinced that it is indeed always possible to impose this kind of conditions. 
Moreover, by proceeding as in the previous section, we deduce that the new values of 
$\hat d^{ABC}$ and $\hat c^A_{\alpha\beta}$ must satisfy the following properties:
\bea
\a\a \hat d^{000} = \text{\small $\frac 2{\sqrt{3}}$} \,,\;\; \hat d^{00a} = 0 \,,\;\; 
\hat d^{0 ab} = - \text{\small $\frac 1{\sqrt{3}}$}\, \delta_{ab} \,,\;\; \hat d^{abc} = \text{generic} \,, \label{dcanhet} \\
\a\a \hat c^0_{\alpha\beta} = \text{\small $\frac 1{\sqrt{3}}$} \, \delta_{\alpha\beta} \,,\;\; 
\hat c^a_{\alpha\beta} = \text{generic} \label{lambdacanhet} \,.
\eea
Notice finally that from the point of view of the Calabi-Yau manifold, the canonical frame just corresponds 
as before to a convenient choice of basis for harmonic forms, which is suitably oriented with respect to 
the K\"ahler form and normalized in such a way as to get unit volume, since 
$\langle \hat v_A \rangle = \sqrt{3}\, \delta_{A0}$ and $\langle \hat {\cal V} \rangle = 1$.
Moreover, by comparing with the general results of section 2 we see that 
the canonical frame again essentially corresponds to choosing a parametrization such that the special direction $k^i$ 
is identified with one of the fields, since $\langle \hat k^i \rangle = \delta^i_0$. 

\subsection{Geometry}

Let us now explore the geometry at a given reference point by using the new canonical coordinates.
For notational simplicity, we drop from now on all the hats referring  to the definition of this special frame, and
also the brackets referring to the special point. 

We start as before by computing the first five partial derivatives of ${\cal V}$, which are the basic ingredients 
that we need. In this case, there is an additional difficulty compared to the previous case, due to the fact that the 
relation between $J^A$ and $v_A$ cannot be explicitly inverted, in general. Fortunately, one can however get 
around this by just using the implicit definition of the $J^A$ in terms of the $v_A$. The Jacobian of this transformation
is $\partial J^A/\partial v_B = d^{AB}$, where $d^{AB} = d^{ABC} v_C$, and its inverse is 
$\partial v_A/\partial J^B = \tilde d_{AB}$, where $\tilde d_{AB}$ is the inverse of the matrix $d^{AB}$. 
Let us also introduce the two new symbols $\tilde d_A = v_A$ and 
$\tilde d_{ABC} = \tilde d_{AE} \tilde d_{BF} \tilde d_{CG} d^{EFG}$. 
These satisfy simple algebraic properties: 
$\tilde d_{AB} J^B = 1/2 \, \tilde d_A$ and $\tilde d_{ABC} J^C = 1/2\, \tilde d_{AB}$. 
Moreover, they depend on $J^A$ but their derivatives with respect to these variables have 
a very simple structure: $\partial \tilde d_{A}/\partial J^B = \tilde d_{AB}$,
$\partial \tilde d_{AB}/\partial J^C = - \tilde d_{ABC}$ and 
$\partial \tilde d_{ABC}/\partial J^D = - \tilde d_{ABF} d^{FG} \tilde d_{GCD} 
- \tilde d_{ADF} d^{FG} \tilde d_{GBC} - \tilde d_{ACF} d^{FG} \tilde d_{GBD}$.
It is then possible to express all derivatives of ${\cal V}$ in terms of the following quantities:
\bea
\a\a \tilde d_A = v_A \,,\;\; d^A = \text{\small $\frac 12$}\, d^{ABC} v_B v_C \,, \\
\a\a \tilde d_{AB} \;\mbox{: inverse of}\; d^{AB} = d^{ABC} v_C \,, \\[1mm]
\a\a \tilde d_{ABC} = \tilde d_{AE} \tilde d_{BF} \tilde d_{CG} d^{EFG} \,.
\eea
After a straightforward computation, one finds that the first five derivatives 
of ${\cal V}$ can be written in the following form:
\bea
\a\a {\cal V}_i = \text{\small $\frac 12$}\, \tilde d_{A} J^A_i \,, \\[0mm]
\a\a {\cal V}_{i \bar \jmath} =  \text{\small $\frac 12$} \tilde d_{AB} J^A_i J^B_{\bar \jmath} 
+ \text{\small $\frac 12$}\, \tilde d_{A} J^A_{i \bar \jmath} \,, \\
\a\a {\cal V}_{i \bar \jmath p} =  - \text{\small $\frac 12$} \tilde d_{ABC} J^A_i J^B_{\bar \jmath} J^C_p 
+ \text{\small $\frac 12$} \tilde d_{AB} \big(J^A_{i \bar \jmath} J^B_p + J^A_{p \bar \jmath} J^B_{i} \big)\,, \\
\a\a {\cal V}_{i \bar \jmath p \bar q} =  \text{\small $\frac 12$} \tilde d_{ABX} d^{XY} \tilde d_{YCD} 
\big(J^A_i J^B_{\bar \jmath} J^C_p J^D_{\bar q} + \mbox{2 p.} \big) \nn \\
\a\a \hspace{35pt}
- \text{\small $\frac 12$} \tilde d_{ABC} \big(J^A_i J^B_{\bar \jmath} J^C_{p \bar q} + \mbox{3 p.} \big) 
+ \text{\small $\frac 12$} \tilde d_{AB} \big(J^A_{i \bar \jmath} J^B_{p \bar q} + J^A_{i \bar q} J^B_{p \bar \jmath} \big) \,, \\
\a\a {\cal V}_{i \bar \jmath p \bar q n}= - \text{\small $\frac 12$} \tilde d_{ABX} d^{XY} \tilde d_{YEZ} d^{ZK} \tilde d_{KCD} 
\big(J^A_{i} J^B_{\bar \jmath} J^C_{p} J^D_{\bar q} J^E_n + \mbox{14 p.}\big) \nn \\
\a\a \hspace{41pt} + \text{\small $\frac 12$} \tilde d_{ABX} d^{XY} \tilde d_{YCD} 
\big(J^A_{i \bar \jmath} J^B_p J^C_{\bar q} J^D_{n\!\!} + \mbox{17 p.} \big) 
- \text{\small $\frac 12$} \tilde d_{ABC} \big(J^A_{i \bar \jmath} J^B_{p \bar q} J^C_{n\!\!} + \mbox{5 p.} \big) \,.
\eea
Using these expressions and going to the canonical frame at the reference point, one then obtains
the following non-vanishing entries for the derivatives of $Y={\cal V}^2$:
\bea
\a\a Y_A = d_A \,, \\[0.5mm]
\a\a Y_{A \bar B} = d_{AB} \,,\;\; Y_{\alpha \bar \beta} = - \delta_{\alpha\beta} \,, \\
\a\a Y_{A \bar B C} = d_{ABC} \,,\;\; Y_{A \alpha \bar \beta} = - d_{AX} c^X_{\alpha\beta} \,, \\[0mm]
\a\a Y_{A \bar B C \bar D} = - 2\, a_{ABCD} ,\; Y_{A \bar B \alpha \bar \beta} = - d_{ABX} c^X_{\alpha\beta} \,,\;
Y_{\alpha \bar \beta \gamma \bar \delta} = d_{XY} \big(c^X_{\alpha \beta} c^Y_{\gamma \delta} 
\!+\! c^X_{\alpha\delta} c^Y_{\gamma\beta}\big) \,, \\[0.5mm]
\a\a Y_{A \bar B C \bar D E} = \big(a_{ABCX} d_{XDE} + \mbox{9 p.}\big) 
+ 2 \big(a_{ABCD} d_E + \mbox{4 p.} \big) \,,\\[0.7mm]
\a\a Y_{A \bar B C \alpha \bar \beta} = 2\, a_{ABCX} c^X_{\alpha\beta} \,,\;\; 
Y_{A \alpha \bar \beta \gamma \bar \delta} = d_{AXY} \big(c^X_{\alpha\beta} c^Y_{\gamma\delta} 
+ c^X_{\alpha\delta} c^Y_{\gamma\beta} \big) \,,
\eea
The parameters in these expressions are now given by 
$d_A = 3/2\, d_{A00}$, $d_{AB} = \sqrt{3}\, d_{AB0}$ and $d_{ABC}$ taking the already 
studied restricted form. To simplify the notation, we have lowered all the indices in these 
quantities with the trivial metric at the reference point. Finally, we introduce as before for convenience 
the following combinations of the parameters $d_{ABC}$ and $c^A_{\alpha\beta}$:
\bea
\a\a a_{ABCD} = \text{\small $\frac 12$} \big(d_{ABX} d_{XCD} \!+ d_{ADX} d_{XBC} \!+ d_{ACX} d_{XBD} \big) \nn \\
\a\a \hspace{48pt} -\, \text{\small $\frac 12$} \big(d_{AB} d_{CD} \!+ d_{AD} d_{BC} \!+ d_{AC} d_{BD} \big) \nn \\
\a\a \hspace{48pt} +\, \text{\small $\frac 12$} \big(d_A d_{BCD} + d_B d_{ACD} + d_C d_{ABD} + d_D d_{ABC} \big) \,, 
\label{aor} \\
\a\a b^{AB}_{\alpha\beta} = \text{\small $\frac 12$} \Big(\big\{c^A,c^B\big\}_{\hspace{-1pt}\alpha\beta} - d_{ABX} c^X_{\alpha\beta} 
+ d_{AB} \delta_{\alpha\beta} - d_A c^B_{\alpha\beta} - d_B c^A_{\alpha\beta}\Big) \,, 
\label{bor} \\
\a\a \tau^A_{\alpha\beta\gamma\delta} = \text{\small $\frac 12$} \Big(\big[c^A,c^X\big]_{\hspace{-1pt}\alpha\beta} c^X_{\gamma\delta} 
\!+\! \big[c^A,c^X\big]_{\hspace{-1pt} \alpha\delta} c^X_{\gamma\beta} 
\!+\! \big[c^A,c^X\big]_{\hspace{-1pt}\gamma\delta} c^X_{\alpha\beta} 
\!+\! \big[c^A,c^X\big]_{\hspace{-1pt} \gamma\beta} c^X_{\alpha\beta} \Big) \,.
\label{tauor}
\eea
These quantities are completely symmetric in their indices of type $A,B,\cdots$ and vanish if one of these
is equal to $0$, meaning that $a_{ABCX} d_X = 0$, $b^{AX}_{\alpha\beta} d_X = 0$ and 
$\tau^X_{\alpha\beta\gamma\delta} d_X = 0$.

Starting from the above expressions, it is now straightforward to compute all the geometric quantities we are interested in.
The metric is trivial and its non-vanishing entries are
\bea
\a\a g_{A \bar B} = \delta_{AB} \,,\\
\a\a g_{\alpha \bar \beta} = \delta_{\alpha\beta} \,.
\eea
The Christoffel connection is instead non-trivial, and its non-vanishing entries are
\bea
\a\a \Gamma_{A B \bar C} = - d_{ABC} + \big(d_{AB} d_C + \mbox{2 p.} \big) - 2\, d_A d_B d_C \,, \\
\a\a \Gamma_{A \alpha \bar \beta} = - c^A_{\alpha\beta} \,.
\eea
The Riemann curvature tensor is found to be given by
\bea
\a\a R_{A \bar B C \bar D} = \delta_{AB} \delta_{CD} + \delta_{AD} \delta_{BC} - d_{ACX} d_{XBD} + 2\, a_{A B C D}\,, \\[0.5mm]
\a\a R_{\alpha \bar \beta \gamma \bar \delta} = c^X_{\alpha\beta} c^X_{\gamma\delta} + c^X_{\alpha\delta} c^X_{\gamma\beta} \,, \\
\a\a R_{\hspace{-1pt} \alpha \bar \beta A \bar B} = \delta_{AB} \delta_{\alpha\beta} 
- d_{AX} d_{BY} (c^X c^Y)_{\alpha\beta} + d_{ABX} c^X_{\alpha\beta} \,.
\eea
Finally, the covariant derivative of the Riemann tensor reads:
\bea
\a\a \nabla_{\!A} R_{B \bar C D \bar E} = -\,a_{ABDX} d_{X C E} 
+ \big(a_{A C E X} d_{X B D} + a_{B C E X} d_{X A D} + a_{D C E X} d_{X A B} \big) \nn \\[1mm]
\a\a \hspace{65pt} -\, \big(a_{ABCX} d_{XDE} + a_{ADCX} d_{XBE} + a_{BDCX} d_{XAE} \nn \\[1mm]
\a\a \hspace{82pt} +\, a_{ABEX} d_{XDC} + a_{ADEX} d_{XBC} + a_{BDEX} d_{XAC} \big) \nn \\[1mm]
\a\a \hspace{65pt} -\,2\, \big(a_{ABDC}d_E + a_{ABDE} d_C \big) \,, \\[1mm]
\a\a \nabla_{\!A} R_{B \bar C \alpha \bar \beta} = -\, a_{ABCX} c^X_{\alpha\beta} + 2\, d_C b^{AB}_{\alpha\beta} 
- d_{ABX} b^{XC}_{\alpha\beta} + d_{ACX} b^{XB}_{\alpha\beta} + d_{BCX} b^{XA}_{\alpha\beta}  \nn \\[1mm]
\a\a \hspace{62pt} -\, \big\{b^{AB}, c^C\big\}_{\hspace{-1pt} \alpha\beta} 
+ \big[b^{AC}, c^B\big]_{\hspace{-1pt} \alpha\beta} + \big[b^{BC}, c^A\big]_{\hspace{-1pt} \alpha\beta}\,, \\[0.5mm]
\a\a \nabla_{\!A} R_{\alpha \bar \beta \gamma \bar \delta} = \tau^A_{\alpha\beta\gamma\delta} 
+ b^{AX}_{\alpha\beta} c^X_{\gamma\delta} + b^{AX}_{\alpha\delta} c^X_{\gamma\beta} 
+ b^{AX}_{\gamma\delta} c^X_{\alpha\beta} + b^{AX}_{\gamma\beta} c^X_{\alpha\delta} \,.
\eea
We thus see that the manifold is a symmetric space with covariantly constant curvature if and only if the quantities 
$a_{ABCD}$, $b^{AB}_{\alpha\beta}$ and $\tau^A_{\alpha\beta\gamma\delta}$ identically vanish, exactly as before.

\section{General features of the geometry}
\setcounter{equation}{0}

From the results of the previous two sections, we discover that the form of the tensors characterizing 
the geometry of the scalar manifolds of heterotic and orientifold models are very similar in the canonical 
frame. This similarity is best and most concisely exhibited by explicitly splitting the indices parallel and orthogonal 
to the special direction defined by the canonical frame: $A=0,a$. The parameters specifying the model
and also the vacuum point are then summarized in the previously defined quantities $d_{abc}$ and $c^a_{\alpha\beta}$,
where we have again dropped the hats for simplicity. It is however useful and convenient to introduce some 
specific notation for various combinations of these parameters, which will turn out to play special roles in the following. 
Recall first that the parameters controlling the deviation from the coset situation have the same expressions 
(\ref{ahet}), (\ref{bhet}), (\ref{tauhet}) and (\ref{aor}), (\ref{bor}), (\ref{tauor}) in both models, and their only 
common non-trivial components are those where all the indices are orthogonal:
\bea
\a\a a_{abcd} = \text{\small $\frac 12$} \big(d_{abr} d_{r c d} \!+ d_{adr} d_{r b c} \!+ d_{acr} d_{r bd} \big)
- \text{\small $\frac 13$} \big(\delta_{ab} \delta_{cd} \!+ \delta_{ad} \delta_{bc} \!+ \delta_{ac} \delta_{bd} \big) \,, \label{a} \\[0mm]
\a\a b^{ab}_{\alpha\beta} = \text{\small $\frac 12$} \big\{c^a, c^b\big\}_{\alpha\beta} 
- \text{\small $\frac 13$} \delta_{ab} \delta_{\alpha\beta} 
- \text{\small $\frac 12$} d_{abr} c^r_{\alpha\beta} \,, \label{b} \\[-0.5mm]
\a\a \tau^a_{\alpha\beta\gamma\delta} = \text{\small $\frac 12$} \Big(\big[c^a, c^r\big]_{\alpha\beta} c^r_{\gamma\delta} 
+ \big[c^a, c^r\big]_{\alpha\delta} c^r_{\gamma\beta} 
+ \big[c^a, c^r\big]_{\gamma\delta} c^r_{\alpha\beta} 
+ \big[c^a, c^r\big]_{\gamma\beta} c^r_{\alpha\delta}\Big) \,. \label{tau}
\eea
Let us next introduce also some short-hand notation for the following additional combinations of parameters, 
which will allow us to write the geometry in a nice and compact form:
\bea
\a\a x_{abcd} = \text{\small $\frac 12$} \Big(d_{abr} d_{rcd} \!+ d_{adr} d_{rbc} \!- d_{acr} d_{rbd}  \big)
+ \text{\small $\frac 23$} \Big(\delta_{ab} \delta_{cd} \!+ \delta_{ad} \delta_{bc} \!- \delta_{ac} \delta_{bd} \Big) \,, \label{x} \\
\a\a y^{ab}_{\alpha\beta} = \text{\small $\frac 12$} \big[c^a, c^b\big]_{\hspace{-1pt}\alpha\beta} 
- \text{\small $\frac 13$} \delta_{ab} \delta_{\alpha\beta} - \text{\small $\frac 12$} d_{abr} c^r_{\alpha\beta} \,. \label{y}
\eea
Finally, in the applications that we will discuss in the last two sections, it will also be useful to define the 
following last couple of quantities:
\bea
\a\a \alpha_{abcd} = - \text{\small $\frac 14$} \Big(d_{abr} d_{rcd} \!+ d_{adr} d_{rbc} \!- 2\, d_{acr} d_{rbd}  \big)
- \text{\small $\frac 13$} \Big(\delta_{ab} \delta_{cd} \!+ \delta_{ad} \delta_{bc} \!-2\, \delta_{ac} \delta_{bd} \Big) \,, \label{alpha} \\
\a\a \beta^{ab}_{\alpha\beta} = \text{\small $\frac 12$} \big[c^a, c^b\big]_{\hspace{-1pt} \alpha\beta} \,. \label{beta}
\eea 

\subsection{Generic case}

Let us first consider generic models with generic values of the parameters $d_{abc}$ and $c^a_{\alpha \beta}$.
These correspond to generic choices of Calabi-Yau manifolds and holomorphic vector bundles over them. With the help 
of the above notation, we can make the results of the previous two sections more explicit and compare them more 
efficiently. The metric is in both cases simply
\bea
\a\a g_{0 \bar 0} = 1 \,,\;\; g_{a \bar b} = \delta_{ab} \,,\\
\a\a g_{\alpha \bar \beta} = \delta_{\alpha\beta} \,.
\eea
The Christoffel connection is also identical in the two cases and given by
\bea
\a\a \Gamma_{00\bar 0} = - \text{\small $\frac 2{\sqrt{3}}$} \,,\;\;
\Gamma_{ab \bar 0} = - \text{\small $\frac 2{\sqrt{3}} \delta_{ab}$} \,,\;\;
\Gamma_{0 a \bar b} = - \text{\small $\frac 2{\sqrt{3}} \delta_{ab}$} \,,\;\;
\Gamma_{ab \bar c} = - d_{abc} \,, \\
\a\a \Gamma_{0 \alpha \bar \beta} = - \text{\small $\frac 1{\sqrt{3}}$} \delta_{\alpha\beta} \,,\;\; 
\Gamma_{a \alpha \bar \beta} = - c^a_{\alpha\beta} \,.
\eea
The Riemann tensor can instead be written in the following simple way, with the upper 
and lower signs applying respectively to heterotic and orientifold models:
\bea
\a\a R_{0 \bar 0 0 \bar 0} = \text{\small $\frac 23$} \,,\;\; R_{0 \bar 0 a \bar b} = \text{\small $\frac 23$}\, \delta_{ab} \,,\;\;  
R_{a \bar 0 b \bar 0} = \text{\small $\frac 23$}\, \delta_{ab} \,,\;\; 
R_{a \bar b c \bar 0} = \text{\small $\frac 1{\sqrt{3}}$}\,  d_{abc} \,, \label{Riemstring1} \\[0.8mm]
\a\a R_{a \bar b c \bar d} = x_{abcd} \mp a_{abcd} \,, \label{Riemstring2} \\[1.2mm]
\a\a R_{\alpha \bar \beta \gamma \bar \delta} = 
\text{\small $\frac 13$} \big(\delta_{\alpha\beta} \delta_{\gamma\delta} + \delta_{\alpha\delta} \delta_{\gamma\beta} \big) 
+ c^r_{\alpha\beta} c^r_{\gamma\delta} + c^r_{\alpha\delta} c^r_{\gamma\beta} \,, \label{Riemstring3} \\
\a\a R_{\hspace{-1pt} \alpha \bar \beta 0 \bar 0} = \text{\small $\frac 13$}\, \delta_{\hspace{-1pt} \alpha\beta} \,,\;\,
R_{\hspace{-1pt} \alpha \bar \beta a \bar b} = - y^{ab}_{\alpha\beta} - b^{ab}_{\alpha\beta}\,,\;\,
R_{\hspace{-1pt} \alpha \bar \beta 0 \bar b} = \text{\small $\frac 1{\!\sqrt{3}}$} c^b_{\hspace{-1pt} \alpha\beta} \,. \label{Riemstring4}
\eea
Finally the covariant derivative of the Riemann tensor also differs only by a few signs for heterotic and orientifold 
models and reads:
\bea
\a\a \nabla_{\!a} R_{b \bar c d \bar 0} = \pm \text{\small $\frac 2{\sqrt{3}}$} a_{abcd} \,, \\[-0.5mm]
\a\a \nabla_{\!a} R_{b \bar c d \bar e} = - \text{\small $\frac {3 \pm 1}2$} \, a_{abdr} d_{rce} 
+ \text{\small $\frac {1 \mp 1}2$} \big(a_{acer} d_{rbd} + 2\; {\rm p.} \big) 
- \text{\small $\frac {1 \mp 1}2$} \big(a_{abcr} d_{rde} + 5 \; {\rm p.} \big) , \\[0mm]
\a\a \nabla_{\!a} R_{b \bar 0 \alpha \bar \beta} = \text{\small $\frac 2{\sqrt{3}}$} b^{ab}_{\alpha\beta} \,, \\[0.2mm]
\a\a \nabla_{\!a} R_{b \bar c \alpha \bar \beta} = \big(\!\pm\! a_{abcr} c^{r\!}
\!-\! d_{abr} b^{rc\!} \!+\! d_{acr} b^{rb\!} \!+\! d_{bcr} b^{ra}\! \!-\! \big\{b^{ab\!}, c^c\big\} 
\!+\! \big[b^{ac\!}, c^b\big] \!+\! \big[b^{bc\!}, c^a\big] \big)_{\hspace{-1pt} \alpha\beta} , \\[2.2mm]
\a\a \nabla_{\!a} R_{\alpha \bar \beta \gamma \bar \delta} = \tau^a_{\alpha\beta\gamma\delta} 
+ b^{ar}_{\alpha\beta} c^r_{\gamma\delta} + b^{ar}_{\alpha\delta} c^r_{\gamma\beta} 
+ b^{ar}_{\gamma\delta} c^r_{\alpha\beta} + b^{ar}_{\gamma\beta} c^r_{\alpha\delta} \,.
\eea
The space is thus generically not symmetric and becomes so if and only if the parameters $d_{abc}$ and $c^a_{\alpha\beta}$ 
are such that $a_{abcd} = 0$, $b^{ab}_{\alpha\beta} = 0$ and $\tau^a_{\alpha\beta\gamma\delta} = 0$.

It is straightforward to show that for a given Calabi-Yau manifold, the scalar manifolds of 
the heterotic and the orientifold models coincide if and only if $a_{abcd} = 0$. 
In such a situation, we see from the formulae derived in previous section that the 
metric, the Christoffel connection, the Riemann tensor and the covariant derivative of the Riemann 
tensor do indeed coincide for the two models.  In fact, one can easily check that in that case the whole
K\"ahler potentials coincide for the two models. Indeed, it can be shown that the condition 
$a_{abcd} = 0$ which is equivalent to $a_{ABCD} = 0$ also implies that 
$d_{XYZ} d_{X(AB}d_{YCD} d_{ZEF)} = 4/3\, d_{(ABC} d_{DEF)}$.
This last relation then directly implies that $Y$ and thus $K$ coincide in the two models, as can be 
seen by comparing (\ref{Yhet}) and the square of (\ref{Vor}) with the relation (\ref{Jvor}).
This result generalizes a similar observation done in \cite{DFT1,DFT2} for models with only moduli fields 
to models involving also matter fields, with the significant difference that the coincidence of the scalar 
manifolds of the two kinds of models no-longer implies that they are symmetric spaces, since 
one may have $b^{ab}_{\alpha\beta} \neq 0$ and/or $\tau^a_{\alpha\beta\gamma\delta} \neq 0$.

\subsection{Coset case}

Let us next discuss the special models with particular values of the parameters $d_{abc}$ and $c^a_{\alpha \beta}$
such that not only $a_{abcd} = 0$ but also $b^{ab}_{\alpha\beta} = 0$ and $\tau^a_{\alpha\beta\gamma\delta} = 0$.
These correspond to very special choices of Calabi-Yau manifold and holomorphic bundle over it, which can for 
instance naturally arise from orbifold constructions. In such a situation, the scalar manifolds of the heterotic and the
orientifold models not only coincide but reduce to a coset manifold. It is therefore of some interest to understand 
when this can occur.
 
The possible solutions to the three equations $a_{abcd} = 0$, $b^{ab}_{\alpha\beta} = 0$ and 
$\tau^a_{\alpha\beta\gamma\delta} = 0$ can in principle be classified, and define a finite list of possibilities for 
such coset spaces, which must be a subclass of all the possible K\"ahler symmetric manifolds described in \cite{CV}. 
For models with only moduli fields, this classification has been explicitly carried out in
\cite{CKVDFdWG,CvP}. The basic observation is that the condition $a_{abcd} = 0$ which can be rewritten 
as $a_{ABCD} = 0$ is also essentially equivalent to the condition $d_{XYE} d_{X(AB}d_{YCD)} = 4/3\, \delta_{E(A} d_{BCD)}$. 
This is easier to study and its solutions were shown to correspond to all the possible special K\"ahler symmetric manifolds. 
For models with also matter fields, a similar classification is presumably possible. The basic missing 
ingredient would be a general study of the other condition $b^{ab}_{\alpha\beta} = 0$, and also 
$\tau^a_{\alpha\beta\gamma\delta} = 0$ although this last condition seems to be most of the times 
automatically satisfied (see the examples below) and we will therefore consider it on a different footing. 
The solutions plausibly provide most of the possible extensions of special K\"ahler symmetric manifolds 
to K\"ahler symmetric manifolds by the addition of matter fields 
besides moduli fields. We will however not attempt here to perform a complete classification. 

It is rather straightforward and instructive to verify that the standard coset scalar manifolds arising in the 
simplest orbifold string models in the untwisted sector do indeed represent non-trivial solutions of the three 
equations $a_{abcd} = 0$, $b^{ab}_{\alpha\beta} = 0$ and $\tau^a_{\alpha\beta\gamma\delta} = 0$. 
To see this, let us assume that the matrices $c^a$ are all traceless and form a compact Lie algebra, 
so that $\big[c^a, c^b] = i f_{abc} c^c$ with completely antisymmetric structure constants $f_{abc}$. 
One then automatically gets $\tau^a_{\alpha\beta\gamma\delta} = 0$. The condition $b^{ab}_{\alpha\beta} = 0$ 
implies instead $\big\{c^a, c^b\} = d_{abc} c^c + 2/3\, \delta_{ab} 1\!\!1$, meaning that $d_{abc}$ is the completely 
symmetric invariant symbol of this algebra. One then finds that ${\rm tr} (c^a c^b) = \kappa\, \delta_{ab}$,
$f_{abc} = - i \kappa^{\text{--}1}{\rm tr}([c^a, c^b] c^c)$ and $d_{abc} = \kappa^{\text{--}1}{\rm tr}(\{c^a, c^b\} c^c)$, 
where $\kappa = {\rm tr} (1\!\! 1)/3$. One finally has to impose the condition $a_{abcd} = 0$, and this dramatically
reduces the possible algebras.  The simplest possibility is the $SU(3)$ algebra generated by the $3 \times 3$ 
matrices $\lambda^a$, and one can then choose $c^a = \lambda^a \otimes 1\!\!1_k$. Other similar solutions can 
then also be obtained by replacing $SU(3)$ with one of its maximal-rank subalgebras $SU(2) \times U(1)$ and 
$U(1) \times U(1)$. In this way, one obtains (see for example \cite{AS1}) the following standard coset no-scale 
manifolds, with $m$ = $8$, $4$ or $2$ non-minimal moduli and $n$ = $3k$, $2k+k'$ or $k+k'+k''$ matter fields:
\bea
\a\a \frac {SU(3,3+k)}{U(1) \times SU(3) \times SU(3+k)} \,, \\
\a\a \frac {SU(2,2+k)}{U(1) \times SU(2) \times SU(2+k)} \times  \frac {SU(1,1+k')}{U(1) \times SU(1+k')} \,, \\
\a\a \frac {SU(1,1+k)}{U(1) \times SU(1+k)} \times  \frac {SU(1,1+k')}{U(1) \times SU(1+k')}  
\times  \frac {SU(1,1+k'')}{U(1) \times SU(1+k'')}  \,.
\eea

To conclude this section, let us discuss the meaning of the parameters $a_{abcd}$, $b^{ab}_{\alpha\beta}$ 
and $\tau^a_{\alpha\beta\gamma\delta}$ in the simplest situations. In the trivial case where there 
is only one modulus and an arbitrary number of matter fields, the above quantities do not exist and 
one always gets a maximally symmetric coset space. The simplest non-trivial case is therefore when there 
are two moduli fields and one matter field, so that all the indices $a,c,\cdots$ and $\alpha,\beta,\cdots$ take 
a single value and can be dropped. In the canonical frame, and denoting for short $d = d_{111}$ and 
$c = c^1_{11}$, the definitions (\ref{a}), (\ref{b}) and (\ref{tau}) then give
\bea
\a\a a = \text{\small $\frac 32$}\, d^2 -1 \,,\;\; 
b = c^2 - \text{\small $\frac 12$} d c - \text{\small $\frac 13$} \,, \;\;
\tau = 0 \,.
\label{coset21}
\eea
Notice also in passing that from (\ref{alpha}) and (\ref{beta}) one gets $\alpha = 0$ and $\beta = 0$. 
One may then wonder whether it is possible to understand in a simple and intuitive way the origin and the 
meaning of the conditions $a=0$ and $b=0$, and perhaps work out their generalization to a generic frame. It turns out that 
this is indeed possible in this simple situation. The basic reason is that there is a unique candidate 
coset space for this type of models, which is $SU(1,2)/(U(1) \times SU(2)) \times SU(1,1)/U(1)$ and can be 
described by a K\"ahler potential of the form $K = - n' \log(T'\! + \bar T'\!  - \Phi' \bar \Phi') - n'' \log(T''\! + \bar T'')$ with
$n' + n'' = 3$. The conditions for getting a coset space in this class of models must then correspond to the conditions 
under which the K\"ahler potential 
$K = - \log \big[1/6\,d_{ABC}J^A J^B J^C\big]$, where $J^A = T^A\! + \bar T^A\! - c^A_{\alpha \beta} \Phi^\alpha \bar \Phi^\beta$ 
with $A,B,\cdots=0,1$ and $\alpha,\beta,\cdots = 1$, takes this simpler form, modulo a field redefinition from 
$T^1$, $T^2$, $\Phi^1$ to $T'$, $T''$, $\Phi'$ and a K\"ahler transformation. It is now straightforward to 
determine under which circumstances this is possible. A first condition is that the cubic polynomial 
defined by the intersection numbers $d_{ABC}$ factorize into two factors. This is possible 
if and only if the discriminant $\Delta$ of this polynomial vanishes, so that there is one real simple root 
$R_1$ and one real double root $R_2$, where:
\bea
\a\a \Delta= - 27 \Big(d_{000}^2 d_{111}^2\! - 3\, d_{001}^2 d_{011}^2\! 
+ 4\, d_{000} d_{011}^3\! + 4\, d_{001}^3 d_{111}\! - 6\, d_{000} d_{001} d_{011} d_{111}\Big) \,, \label{Delta}\\
\a\a R_1 = - \frac {d_{001}}{d_{000}} - \frac 2{d_{000}}\,\, \raisebox{9pt}{\footnotesize $3$}\!\!\!\!
\sqrt{d_{001}^3\! - \text{\small $\frac 32$} \, d_{000} d_{001} d_{011}\! + \text{\small $\frac 12$}\, d_{000}^2 d_{111}} \,, \\
\a\a R_2 = - \frac {d_{001}}{d_{000}} + \frac 1{d_{000}}\,\, \raisebox{9pt}{\footnotesize $3$}\!\!\!\!
\sqrt{d_{001}^3\! - \text{\small $\frac 32$} \, d_{000} d_{001} d_{011}\! + \text{\small $\frac 12$}\, d_{000}^2 d_{111}} \,.
\eea
In that case the K\"ahler potential factorizes into the sum of one trivial and two non-trivial pieces:
$K = - \log \big[1/6\, d_{000}] - \log \big[J^0\!-\! R_1 J^1\big] - 2 \log \big[J^0 \!-\! R_2 J^1\! \big]$. 
A second condition is then that the matter fields appear either in the second or the third term but 
not simultaneously in both. It is straightforward to check that this requires that either 
$c^0_{11} - R_1 c^1_{11}$ or $c^0_{11} - R_2 c^1_{11}$ vanishes, or equivalently that their 
product vanishes. To sum up, the two conditions for the space 
to degenerate into a coset are in this case:
\bea
\a\a \Delta = 0 \,, \label{condgen1}\\
\a\a (c^0_{11}\!-\! R_1 c^1_{11})(c^0_{11}\!-\! R_2 c^1_{11}) = 0 \,. \label{condgen2}
\eea
It is now straightforward to verify that in the canonical frame one has $\Delta = - 24\, a$, and that 
whenever $a = 0$ one finds $(c^0_{11}\!-\! R_1 c^1_{11})(c^0_{11}\!-\! R_2 c^1_{11}) = - b$. This 
shows that the combination of the two conditions (\ref{condgen1}) and (\ref{condgen2}) is equivalent 
to the combination of the conditions $a = 0$ and $b = 0$ in the canonical frame, and evidently represents 
their generalization to arbitrary frames. From the above reasoning, it is however also clear that in the more general 
case where more than two moduli fields and/or more than one matter field are present, the situation is much 
more complicated to study from this frame-independent perspective. On the other hand, the conditions in the canonical frame
simply generalize to the conditions $a_{abcd} = 0$, $b^{ab}_{\alpha \beta} = 0$ and $\tau^a_{\alpha \beta \gamma \delta} = 0$.

\section{SGoldstino mass and vacuum metastability}
\setcounter{equation}{0}

As a first application of the results derived in the previous sections, let us consider the condition for 
the existence of a metastable supersymmetry breaking vacuum. This is controlled by the sign of 
the average sGoldstino square mass and depends on the holomorphic sectional curvature of the 
scalar manifold along the Goldstino direction $f^I$. More precisely, assuming for simplicity a negligibly 
small cosmological constant, the average sGoldstino mass is given by 
\be
m^2_{f \bar f} = 3 \Big(R(f) + \text{\small $\frac 23$} \Big) \, m_{3/2}^2 \,,
\ee
where the holomorphic sectional curvature $R(f)$ is defined as
\be
R(f) = - R_{I \bar J K \bar L} f^I \hspace{-1pt} \bar f^{\bar J} \! f^{\hspace{-1pt} K} \hspace{-3pt} \bar f^{\bar L} \,,
\ee
and the vector $f^I$ is subject to the following constraint:
\be
|f|^2 = g_{I \bar J} f^I \! \bar f^{\bar J} = 1 \,.
\ee
A necessary condition for metastability is that $m^2{\!\!}_{f \bar f} > 0$, which implies $R(f) > -2/3$. 
This condition becomes also sufficient whenever the superpotential can be arbitrarily tuned, 
and the upper bound represented by $m^2{\!\!}_{f \bar f}$ on the square mass of the lightest particle can then 
be saturated. In the presence of a positive cosmological constant $V$ parametrized by 
$\gamma = V/(3\,m_{3/2}^2)$ this bound becomes stronger and reads 
$R(f) > -2/3 \, (1 + \gamma)^{-1}$ \cite{CGGLPS2}. 
The effect of vector multiplets has instead been studied in \cite{GRS3}, and it has also been pointed 
out in \cite{BS} that in the presence of broken gauge symmetries the lightest scalar is in fact a 
combination of the sGoldstino and the complex partners of the Goldstones.

In the class of models that we considered, the hidden sector triggering supersymmetry breaking 
can involve both the dilaton $S$ and a subset of the K\"ahler moduli and matter fields $Z^i = T^A,\Phi^\alpha$. 
We can thus have $f^S \neq 0$ and $f^i \neq 0$. Since the dilaton sector and the 
K\"ahler moduli plus matter field sector are factorized, it is convenient to introduce an angle $\theta$ 
to explicitly parametrize the splitting of the Goldstino direction along the two corresponding 
submanifolds and rewrite $f^S = \sin \theta \, g^S$ and $f^i = \cos \theta \, h^i$, where now $|g| = 1$
and $|h|=1$. 
We will imagine here that such a direction can a priori be arbitrary, as in \cite{KL,BIM} (see also \cite{KM}),
and shall not discuss the possibilities offered by specific effects like classical fluxes or 
non-perturbative quantum corrections (see for example \cite{Softflux1,Softflux2,Softflux3,Softflux4})
for recent studies on this).
Recalling that the sectional curvature of the fixed coset manifold $SU(1,1)/U(1)$ describing 
the dilaton is constant and equal to $R(g) = -2$, and parametrizing the sectional curvature of the generic no-scale manifold 
${\cal M}_{Y,N}$ describing the K\"ahler moduli and matter fields as $R(h) = -2/3 + \Sigma(h)$, one can then 
write $R(f)$ in the following form:
\bea
R(f) = -2 \sin^4\! \theta + \Big(\!-\! \text{\small $\frac 23$} + \Sigma(h) \Big) \cos^4\! \theta \,.
\eea
The average sGoldstino mass is correspondingly written as
\bea
m^2_{f \bar f} = \Big[\!-\! 4\, \sin^4\! \theta + 4\, \sin^2\! \theta \cos^2\! \theta + 3\, \Sigma(h) \cos^4\! \theta\Big] m^2_{3/2} \,.
\label{mffcan}
\eea
The quantity $\Sigma(h)$ can be non-zero only if the no-scale manifold ${\cal M}_{Y,N}$ differs from the 
minimal possibility $SU(1,1+n)/(U(1) \times SU(n))$. It measures the amount by which the sectional 
curvature deviates from the critical value $-2/3$, and controls therefore the possibility of making 
$m^2_{f \bar f} \neq 0$ even when $\theta = 0$. A quite explicit but still general expression
for it can be derived by using the general properties of the geometry of no-scale manifolds derived in 
section 2, with $Y$ homogeneous of degree three in $J^A$ and $N^A$ function of $\Phi^\alpha \bar \Phi^\beta$, 
under the simplifying assumption that the matter fields take vanishing expectation values.
Using the short-hand notation in which at the considered point the moduli index $A$ is split into the 
values $0$ corresponding to the direction parallel to $k^A$ and the values $a$ corresponding to the 
directions orthogonal to $k^A$, and reading off the values of the metric, the Christoffel symbol and the Riemann
tensor with at least one parallel index from eqs.~(\ref{genbegin})--(\ref{genend}), one then finds:
\bea
\Sigma(h) \b=\b A_{a \bar b c \bar d}\, h^a \bar h^{\bar b} h^c \bar h^{\bar d} 
+ 4\, B_{a \bar b \alpha \bar \beta}\, h^a \bar h^{\bar b} h^\alpha \bar h^{\bar \beta} 
+ E_{\alpha \bar \beta \gamma \bar \delta}\, h^\alpha \bar h^{\bar \beta} h^\gamma \bar h^{\bar \delta}
- 2\, S^r(h) S_r(h) \,. \nn
\eea
where 
\bea
\a\a A_{a \bar b c \bar d} = \text{\small $\frac 13$} \big(g_{a \bar b} g_{c \bar d} + g_{a \bar d} g_{c \bar b} \big)
- R_{a \bar b c \bar d} 
+ \text{\small $\frac 14$} \big(\Gamma_{r a \bar b} g^{r \bar s} \Gamma_{\bar s \bar d c} 
+ \Gamma_{r a \bar d} g^{r \bar s} \Gamma_{\bar s \bar b c} \big) 
\,, \label{A} \\
\a\a B_{a \bar b \alpha \bar \beta} = \text{\small $\frac 13$} g_{a \bar b} g_{\alpha \bar \beta}
- R_{a \bar b \alpha \bar \beta} 
+ \text{\small $\frac 12$} \Gamma_{r a \bar b}g^{r \bar s} \Gamma_{\bar s \bar \beta \alpha} 
\,, \label{B} \\
\a\a E_{\alpha \bar \beta \gamma \bar \delta} = \text{\small $\frac 13$} \big(g_{\alpha \bar \beta} g_{\gamma \bar \delta}
+ g_{\alpha \bar \delta} g_{\gamma \bar \beta} \big) - R_{\alpha \bar \beta \gamma \bar \delta} 
+ \Gamma_{r \alpha \bar \beta} g^{r \bar s} \Gamma_{\bar s \bar \delta \gamma} 
+ \Gamma_{r \alpha \bar \delta} g^{r \bar s} \Gamma_{\bar s \bar \beta \gamma} 
\label{E} \,,
\eea
and 
\bea
\a\a S_r(h) = \text{\small $\frac 1{\sqrt{3}}$} \big(h_{\bar r} \bar h^{\bar 0} + \bar h_r h^0 \big)
- \text{\small $\frac 12$} \big(\Gamma_{r a \bar b} h^a \bar h^{\bar b} 
+ 2\, \Gamma_{r \alpha \bar \beta} h^\alpha \bar h^{\bar \beta}\big) \,.
\eea
The explicit form of the normalization condition for $h$ is:
\be
|h^0|^2 + g_{a \bar b} h^a \bar h^{\bar b} + g_{\alpha \bar \beta} h^\alpha \bar h^{\bar \beta} = 1 \,.
\label{norm}
\ee
Using the same strategy as in \cite{CGGLPS1}, we now observe that a simple bound on $\Sigma(h)$ can be obtained by 
dropping the sum of squares in the last term, which give negative-definite contributions, and keeping 
the first three terms, which have a priori indefinite signs. A necessary condition for the existence of any 
direction $h$ along which $\Sigma(h)$ is larger than $0$ is then that the sum of these first three terms 
be larger than $0$ for some $h$. In fact, the maximal value $\Sigma_{\rm up}$ of the sum of these three 
terms represents an upper bound on how big the full $\Sigma(h)$ can be, and thus on its maximum $\Sigma_{\rm max}$. 
We thus deduce that 
\be
\Sigma(h) \le \Sigma_{\rm max} \le \Sigma_{\rm up} \,, 
\ee
where:
\be
\Sigma_{\rm up} = \max_{h} \Big\{A_{a \bar b c \bar d}\, h^a \bar h^{\bar b} h^c \bar h^{\bar d} 
+ 4\, B_{a \bar b \alpha \bar \beta}\, h^a \bar h^{\bar b} h^\alpha \bar h^{\bar \beta} 
+ E_{\alpha \bar \beta \gamma \bar \delta}\, h^\alpha \bar h^{\bar \beta} h^\gamma \bar h^{\bar \delta} \Big\} \,.
\ee

We now want to evaluate more explicitly the quantities $\Sigma(h)$ and $\Sigma_{\rm up}$ in the specific cases of Calabi-Yau 
string models of the heterotic and orientifold types, where the Riemann tensor and the Christoffel connection
have a more constrained form parametrized in terms of some numbers $d_{ABC}$ and $c^A_{\alpha \beta}$.
To do so, it is very convenient to go to the canonical frame defined in sections 3 and 4. In this way, one 
can use the simple characterization of the geometry derived in section 5, and after a straightforward computation 
one finds that the quantities (\ref{A}), (\ref{B}) and (\ref{E}) reduce to the following combinations of the quantities 
(\ref{a}), (\ref{b}), (\ref{alpha}) and (\ref{beta}):
\bea
\a\a A_{a \bar b c \bar d} = \pm a_{abcd} + \alpha_{abcd} \,,\;\; 
B_{a \bar b \alpha \bar \beta} = b^{ab}_{\alpha\beta} + \beta^{ab}_{\alpha\beta} \,,\;\;
E_{\alpha \bar \beta \gamma \bar \delta} = 0 \,.
\eea
One then finds
\bea
\Sigma(h) \b=\b \big(\!\pm a_{abcd} + \alpha_{abcd} \big) h^a \bar h^{\bar b} h^c \bar h^{\bar d} 
+ 4 \big(b^{ab}_{\alpha\beta} + \beta^{ab}_{\alpha\beta} \big) h^a \bar h^{\bar b} h^\alpha \bar h^{\bar \beta} \nn \\
\b\;\b - 2\, {\sum}_a \bigg[\text{\small $\frac 1{\sqrt{3}}$} \Big(\bar h^{\bar 0} h^a + h^0 \bar h^{\bar a}\Big) + \text{\small $\frac 12$} 
\Big(d_{abc} h^b \bar h^{\bar c} + 2\, c^a_{\alpha\beta} h^\alpha \bar h^{\bar \beta} \Big)\bigg]^2 \,,
\label{Sigmacan}
\eea
and thus
\be
\Sigma_{\rm up} = \max_{h} \Big\{\big(\!\pm a_{abcd} + \alpha_{abcd} \big) h^a \bar h^{\bar b} h^c \bar h^{\bar d} 
+ 4 \big(b^{ab}_{\alpha\beta} + \beta^{ab}_{\alpha\beta} \big) h^a \bar h^{\bar b} h^\alpha \bar h^{\bar \beta} \Big\} \,.
\label{Sigmaupcan}
\ee
We see that the structure of $\Sigma(h)$ is very similar in heterotic and orientifold models, the only difference 
being the sign with which the $a$ parameter related to moduli enters, as already noticed in 
\cite{CGGLPS1}, while the $b$ parameter related to matter fields enters with the same sign. The average sGoldstino 
mass correspondingly also takes very similar forms. We further notice that $\Sigma(h)$ has a very simple 
dependence on $h^0$, while the functional defining $\Sigma_{\rm up}$ does not depend at all on $h^0$.
This results in two distinct behaviors for 
directions $h^i$ that are parallel and orthogonal to $k^i$. In the parallel direction defined by taking $h^0 =1$ 
and $h^a,h^\alpha = 0$, one finds a trivially vanishing $\Sigma(h)$. In the orthogonal directions defined by 
taking $h^0 = 0$ and $h^a, h^\alpha \neq 0$, one instead 
finds a generically non-trivial and potentially positive $\Sigma(h)$. Notice also that in hybrid directions 
where $h^a = 0$ and $h^0, h^\alpha \neq 0$, one finds again 
a vanishing $\Sigma(h)$ if the further constraints $c^a_{\alpha \beta} h^\alpha \bar h^{\bar \beta} = 0$ hold true.
In such a situation, the sGoldstino mass would then be given by the following bounded expression:
\be
m^2{\!\!}_{f \bar f} = \Big[\!-\! 4 \sin^4\! \theta + 4\,\sin^2\! \theta \cos^2\! \theta\Big]\, m^2_{3/2} 
\;\;\text{if}\;\; h^a = c^a_{\alpha \beta} h^\alpha \bar h^{\bar \beta} = 0 \,.
\label{mffspecial}
\ee
The above remarks also show that $\Sigma_{\rm up} \ge 0$, because the functional involved in the expression
(\ref{Sigmaupcan}) always takes a vanishing value along the parallel and the hybrid directions and 
possibly a positive value along some orthogonal directions. 

\subsection{Necessary conditions for metastability}

A first non-trivial question about $\Sigma(h)$ is to determine whether it can be positive, since 
this would allow for the existence of metastable supersymmetry breaking vacua even when the 
dilaton does not contribute to supersymmetry breaking. In the hope of finding some simple 
necessary conditions for this, one may then try to compute the sign of the associated $\Sigma_{\rm up}$ 
and determine for which models it can be positive. To proceed, we parametrize 
the complex Goldstino variables in terms of a modulus and a phase, as $h^i = \tilde h^i e^{i \delta^i}\!$. 
We then notice that when all the phases vanish, the terms involving the quantities $\alpha_{abcd}$ and 
$\beta^{ab}_{\alpha \beta}$ 
drop out, and the functional problem substantially simplifies. Let us then restrict to the situation where 
we deliberately fix $\delta^i = 0$ and optimize only with respect 
to the real variables $\tilde h^i$, subject to the constraint 
$\sum_i (\tilde h^i)^2 = 1$. This defines the new  quantity
\bea
\a\a \tilde \Sigma_{\rm up} = \max_{\tilde h}
\Big\{\!\pm a_{abcd}\, \tilde h^a \tilde h^b \tilde h^c \tilde h^d 
+ 4\,b^{ab}_{\alpha\beta}\, \tilde h^a \tilde h^b \tilde h^\alpha \tilde h^\beta \Big\} \,.
\eea
Clearly the extremum $\tilde \Sigma_{\rm up}$ of the simplified functional for real $\tilde h^i$ will in general 
be smaller than the extremum $\Sigma_{\rm up}$ of the full original functional for complex $h^i$. But it is 
legitimate to hope that the gross behavior of $\tilde \Sigma_{\rm up}$ as a function of the parameters of 
the models could still give a reasonable indication of the gross behavior of $\Sigma_{\rm up}$. We have 
numerically checked in some simple classes of models that this is indeed the case. We will therefore perform an 
analytical study of the properties of the simplified $\tilde \Sigma_{\rm up}$ and then give a qualitative 
discussion of the properties of the true $\Sigma_{\rm up}$.

To discuss the maximization defining $\tilde \Sigma_{\rm up}$, it is convenient to introduce two angles 
$\chi$ and $\xi$ to explicitly parametrize the distribution of the real Goldstino direction $\tilde h^i$ along the three 
different types of fields arising in the no-scale sector and rewrite $\tilde h^0 = \cos \chi \, \tilde x^0$, 
$\tilde h^a = \sin \chi \cos \xi\, \tilde y^a$ and $\tilde h^\alpha = \sin \chi \sin \xi\, \tilde z^\alpha$, where now 
$|\tilde x| = 1$, $|\tilde y| = 1$ and $|\tilde z| = 1$.
In this way, one can rewrite
\be
\tilde \Sigma_{\rm up} = \max_{\chi,\xi,\tilde y,\tilde z} \Big\{\sin^4 \! \chi 
\Big(\!\pm a(\tilde y) \cos^4\! \xi + 4\, b(\tilde y,\tilde z) \cos^2\! \xi \sin^2\! \xi \Big) \Big\} \,,
\ee
where
\bea
a(\tilde y) = a_{abcd}\, \tilde y^a \tilde y^b \tilde y^c \tilde y^d \,,\;\;
b(\tilde y,\tilde z) = b^{ab}_{\alpha \beta}\, \tilde y^a \tilde y^b \tilde z^\alpha \tilde z^\beta \,.
\eea
It becomes now obvious that $\tilde \Sigma_{\rm up}$ can be positive only if $\pm a(\tilde y)$ or $b(\tilde y,\tilde z)$ 
can be positive. We therefore conclude that:
\be
\tilde \Sigma_{\rm up} > 0 \;\;\text{requires}\;\; 
(\pm a)_{\rm up} \equiv \max_{\tilde y} \big\{\!\pm\! a(\tilde y)\big\} > 0 \;\;\text{or}\;\; 
b_{\rm up} \equiv \max_{\tilde y, \tilde z} \big\{b(\tilde y, \tilde z)\big\} > 0 \,.
\ee 
For fixed $\tilde y^a$ and $\tilde z^\alpha$ and thus $a$ and $b$, the extrema of the functionals $F(\chi) = \sin^4 \! \chi$
and $G(\xi) = \pm\, a \cos^4\! \xi + 4\, b \cos^2\! \xi \sin^2\! \xi$ are straightforward to find. For $F(\chi)$, the are always 
two extrema: the first is at $\chi = 0$ and gives $F = 0$, the second is at $\chi = \pi/2$ and gives $F = 1$.
For $G(\xi)$ there are in general three extrema: the first is at $\xi = 0$ and gives $G = \pm a$, the second is 
at $\xi = \pi/2$ and gives $G=0$, the third is at $\chi = \arccos \sqrt{2\,b/(4\,b\mp a)}$ 
and leads to $G = 4\, b^2/(4\,b\mp a)$. Notice however that while the first and the second of these always exist,
the third exists when $b \not \in\, ]\min\{0, \pm a/2\},\max\{0, \pm a/2\}[\!$ implying that $4\, b^2/(4\,b\mp a) \le \max\{0,2\,b\}$
but not when $b \in\, ]\min\{0, \pm a/2\},\max\{0, \pm a/2\}[\!$ implying anyhow that $4\, b^2/(4\,b\mp a) \le \max\{0,\pm a\}$.
These results confirm that the maximum of the product functional $F(\chi) G(\xi)$ can be positive only if $a > 0$ or $b > 0$, while 
it is zero if $a \le 0$ and $b \le 0$. Moreover, they allow to compute the precise value of $\tilde \Sigma_{\rm up}$ and to 
derive a simple bound on it. With an obvious notation one finds:
\bea
\tilde \Sigma_{\rm up} \b=\b \max \Big\{0, (\pm a)_{\rm up}, \Big(\frac {4\,b^2}{4\, b \mp a}\Big)_{\!\! \rm up}\Big\}  \nn \\
\b\le\b \max \Big\{0, (\pm a)_{\rm up}, 2\,b_{\rm up}\Big\} \,.
\label{condreal}
\eea
This concludes our analytic study of the necessary conditions for metastability in the approximation 
in which the sGoldstino direction is assumed to be real and without relying on the dilaton. The result 
(\ref{condreal}) shows that for coset spaces where $a_{abcd} = 0$ and $b^{ab}_{\alpha \beta} = 0$ 
one always finds $\tilde \Sigma_{\rm up} = 0$ since $(\pm a)_{\rm up} = 0$ and $b_{\rm up} = 0$, while 
for non-coset spaces where $a_{abcd} \neq 0$ or $b^{ab}_{\alpha \beta} \neq 0$ one can obtain 
$\tilde \Sigma_{\rm up} > 0$ only if $(\pm a)_{\rm up} > 0$ or $b_{\rm up} > 0$. This generalizes 
the result that was derived in \cite{CGGLPS1} (see also  \cite{CGGLPS2,CGGPS}) for models with 
two moduli and no matter fields to models with an arbitrary number of moduli and matter fields.

The maximization defining the true $\Sigma_{\rm up}$ is more complicated, but as already anticipated its 
behavior is qualitatively similar to that of $\tilde \Sigma_{\rm up}$, as a matter of fact. The 
main difference is that the necessary condition to get $\Sigma_{\rm up} > 0$ is not exactly given by a sharp 
conditions related just to $a_{abcd}$ and $b^{ab}\hspace{-9.5pt} \raisebox{-3pt}{\scriptsize $\alpha \beta$}$, 
but is slightly affected also by $\alpha_{abcd}$ and $\beta^{ab}\hspace{-9.5pt} \raisebox{-3pt}{\scriptsize $\alpha \beta$}$. 
More precisely, one finds a slight blurring of the sharp conditions 
that were required for $\tilde \Sigma_{\rm up} >0$, and the requirements $(\pm a)_{\rm up} \gsim 0$ or $b_{\rm up} \gsim 0$
are now only qualitatively true far away from the transition points where $(\pm a)_{\rm up} \sim 0$ or $b_{\rm up} \sim 0$.
This blurring is however quite limited and not always efficient. For instance, in the particular case of coset spaces, 
where $a_{abcd} = 0$ and $b^{ab}\hspace{-9.5pt} \raisebox{-3pt}{\scriptsize $\alpha \beta$} = 0$ 
but in general $\alpha_{abcd} \neq 0$ and 
$\beta^{ab}\hspace{-9.5pt} \raisebox{-3pt}{\scriptsize $\alpha \beta$} \neq 0$, one manifestly has $\tilde \Sigma_{\rm up} = 0$, 
but as a matter of fact one can verify case by case that one also finds $\Sigma_{\rm up} = 0$ and in fact $\Sigma_{\rm max} = 0$, 
as was first derived in \cite{GRS2} for a large class of examples. In that case, we therefore see that the absence 
of the terms involving $a_{abcd}$ and $b^{ab}\hspace{-9.5pt} \raisebox{-3pt}{\scriptsize $\alpha \beta$}$ is the 
crucial feature, while the presence of the terms involving $\alpha_{abcd}$ and 
$\beta^{ab}\hspace{-9.5pt} \raisebox{-3pt}{\scriptsize $\alpha \beta$}$ is essentially irrelevant, in the sense that the maximal 
value of $\Sigma_{\rm up}$ is realized along directions such that these terms vanish. More in general, we found evidence
through numerical investigations that also for generic non-coset spaces the crucial features are again 
controlled by $a_{abcd}$ and $b^{ab}\hspace{-9.5pt} \raisebox{-3pt}{\scriptsize $\alpha \beta$}$, while the presence of 
the terms involving $\alpha_{abcd}$ and $\beta^{ab}\hspace{-9.5pt} \raisebox{-3pt}{\scriptsize $\alpha \beta$}$
gives only small effects. To sum up, this leads us to argue that if neither $(\pm a)_{\rm up}$ nor $b_{\rm up}$ are positive 
then the average sGoldstino mass is essentially bounded:
\be
m^2_{f \bar f} \lsim \Big[\!-\! 4\, \sin^4\! \theta + 4 \sin^2\! \theta\, \cos^2\! \theta\Big] \, m_{3/2}^2 \;\;\text{when}\;\;
(\pm a)_{\rm up}, b_{\rm up} \lsim 0 \,.
\ee
In particular, in such a situation one would get $m^2_{f \bar f} \lsim - 4\,m_{3/2}^2$ in the dilaton dominated case,
$m^2_{f \bar f} \lsim 0$ in the moduli dominated case, and $m^2_{f \bar f} \lsim 1/2 \,m_{3/2}^2$ in any case.

Let us finally discuss the meaning of the signs of $(\pm a)_{\rm up}$ and $b_{\rm up}$ in the generic frame in which 
the model is defined. Recall first that the quantities $d_{abc}$ and $c^a\hspace{-5pt} \raisebox{-3pt}{\scriptsize $\alpha \beta$}$ 
defined in the canonical frame implicitly depend on the original parameters $d_{ABC}$ and 
$c^A\hspace{-6.5pt} \raisebox{-3pt}{\scriptsize $\alpha \beta$}$ as 
well as on the reference point $P$. The same is therefore true also for the quantities $\pm a_{abcd}$ and 
$b^{ab}\hspace{-9.5pt} \raisebox{-3pt}{\scriptsize $\alpha \beta$}$ as well as their extrema $(\pm a)_{\rm up}$ and 
$b_{\rm up}$. A crucial question is then whether the signs of $(\pm a)_{\rm up}$ and $b_{\rm up}$ 
are fixed within a given model specified by a choice of parameters 
$d_{ABC}$ and $c^A\hspace{-6.5pt} \raisebox{-3pt}{\scriptsize $\alpha \beta$}$ or can instead be changed by 
changing the point $P$ within the given model. To try to answer this question, we notice that the generalizations of the 
quantities $\pm a_{abcd}$ and $b^{ab}\hspace{-9.5pt} \raisebox{-3pt}{\scriptsize $\alpha \beta$}$ defined in the 
canonical frame to an arbitrary frame are essentially given by the quantities $A \raisebox{-2pt}{\scriptsize $a \bar b c \bar d$}$ and 
$B\raisebox{-2pt}{\scriptsize $a \bar b \alpha \bar \beta$}$ defined in (\ref{A}) and (\ref{B}), at least if one ignores the 
effects of the terms involving $\alpha_{abcd}$ and $\beta^{ab}\hspace{-9.5pt} \raisebox{-3pt}{\scriptsize $\alpha \beta$}$. 
This shows that the behavior of $\pm a_{abcd}$ is sensitive to moduli but not matter fields, while the behavior of 
$b^{ab}\hspace{-9.5pt} \raisebox{-3pt}{\scriptsize $\alpha \beta$}$ depends both on moduli and matter fields.
One can then try to evaluate more explicitly these expressions to understand how they are allowed to depend on $P$. 
In the simplest non-trivial case of models with two moduli fields and one matter field, this can indeed be done rather explicitly
thanks to the fact that all the indices can take a single value and can thus be dropped. One is then left with just two 
parameters $A$ and $B$ which are in one-to-one correspondence with the parameters $\pm a$ and $b$ controlling 
the deviations from the coset space situation (since in this case $\tau = 0$), and this allows to argue quite sharply about 
what kind of dependence 
on $P$ is allowed. We first notice that the sign of $A$ cannot change when continuously changing $P$. 
Indeed, if $A = 0$ for some $P$ then it is effectively as if one had a coset space, since when studying $A$ 
one can focus on moduli fields and completely ignore matter fields so that the value of $B$ does not matter, 
and one must then get $A = 0$ at any other $P$ too. This statement can be explicitly checked. Indeed, a 
straightforward computation shows that $A = - \Delta/24\, e^{4 K}\! \det^{-3}\! g$, where $\Delta$ is 
given by (\ref{Delta}) and the remaining factor depends on $P$ but is not allowed to change sign. 
We next notice that the sign of $B$ can instead change when continuously 
changing $P$, because even if $B=0$ at some point $P$ one does not necessarily have a coset space, since 
when studying $B$ one must consider both the moduli and the matter fields so that the value of $A$ matters. 
If however one starts from a situation where $A = 0$, then even $B$ is no longer allowed to change sign by 
continuously changing $P$, because if $B = 0$ for some $P$ one has a coset space and one must then have 
$B = 0$ also at any other $P$. These statements can be verified numerically, but we were not able to find any 
simple universal expression for $B$ that could make them manifest. In more general situations with more than 
two moduli fields and/or more than 
one matter field, the situation is clearly more complicated, since there are more parameters. It is then a priori always 
possible that $A(y)$ and $B(y,z)$ change sign when changing continuously $P$, because this does not imply
that all the components of $A\raisebox{-2pt}{\scriptsize $a \bar b c \bar d$}$ and 
$B \raisebox{-2pt}{\scriptsize $a \bar b \alpha \bar \beta$}$ go through zero simultaneously. In other words,
in this more general case the coset space situations do no longer separate the parameter space into
semi-disconnected parts.

\subsection{Upper bound on the mass of the lightest scalar}

A second non-trivial question about $\Sigma(h)$ is to compute the maximal value $\Sigma_{\rm max}$ that it is  
allowed to take, since this allows to set an upper bound on the mass of the lightest scalar relative to $m_{3/2}$ which 
can have relevant cosmological implications (see for instance \cite{CGGLPS2,AKK}). To facilitate the discussion, we again 
introduce two angles $\chi$ and $\xi$ and parametrize the complex Goldstino direction $h^i$ in the usual form 
$h^0 = \cos \chi \, x^0$, $h^a = \sin \chi \cos \xi\, y^a$ and $h^\alpha = \sin \chi \sin \xi\, z^\alpha$, where $|x| = 1$, $|y| = 1$ 
and $|z| = 1$. 

A general preliminary information that can be easily extracted concerns the absolute maximum that can be achieved 
for $\Sigma(h)$ within each class of models by suitably dialing not only the Goldstino direction $h^i$ but 
also the parameters $d_{abc}$ and $c^a_{\alpha \beta}$. To derive such an absolute bound, we note that 
from the definitions of $a_{abcd}$, $\alpha_{abcd}$, $b^{ab}_{\alpha \beta}$ and $\beta^{ab}_{\alpha \beta}$ 
it follows that in eq.~(\ref{Sigmacan}) the first term involving $\pm a_{abcd} + \alpha_{abcd}$ can 
be arbitrarily large in the heterotic case but at most unity in the orientifold case, while the second term involving 
$b^{ab}_{\alpha \beta} +\beta^{ab}_{\alpha \beta}$  can be arbitrarily large in both cases. This means that
when only moduli fields participate in supersymmetry breaking one gets $\Sigma(h) < + \infty$ for heterotic 
models but $\Sigma(h) < 1$ for orientifold models. On the other hand, when also matter fields participate in 
supersymmetry breaking one gets $\Sigma(h) < +\infty $ both for heterotic and orientifold models, and the 
situation therefore significantly improves. These extreme values of 
$\Sigma(h)$ can be obtained when the parameters $d_{abc}$ and $c^a_{\alpha \beta}$ are either
very large or very small, implying that $a_{abcd}$ and $b^{ab}_{\alpha \beta}$ are necessarily 
non-zero and the model is thus far away from any coset. By studying these limits one can 
then determine more explicitly the behavior of $\Sigma(h)$ and its maximum 
$\Sigma_{\rm max}$ in these asymptotic regions. 

Let us first consider the case where some of the parameters $d_{abc}$ and $c^a_{\alpha \beta}$ are large.
In such a situation one may keep only those terms in (\ref{Sigmacan}) that involve two powers of the
parameters $d_{abc}$ and $c^a_{\alpha \beta}$. This leads to the following expression:
\bea
\Sigma(\chi,\xi,y,z) \b\simeq\b \sin^4\! \chi \bigg[{\sum}_r \Big(\!-(1\mp 1) \, (d_r(y))^2
+ \text{\small $\frac {1 \pm 1}2$} \, |\hat d_r(y)|^2 \Big)\cos^4\! \xi \nn \\[0mm]
\b\;\b \hspace{30pt} +\,4\, \Big({\sum}_\epsilon |\hat c_\epsilon(y,z)|^2 - {\sum}_r d_r(y) c^r(z)\Big) 
\cos^2\! \xi \sin^2\! \xi \nn \\[0mm]
\b\;\b \hspace{30pt} -\, 2\, {\sum}_r (c^r(z))^2 \sin^4\! \xi \bigg] \,,
\label{Sigmalargedc}
\eea
where
\bea
\a\a d_r(y) = d_{rab} y^a \bar y^{\bar b} \,,\;\; 
\hat d_r(y) = d_{rab} y^a y^b \,, \\
\a\a c^r(z) = c^r_{\alpha \beta} z^\alpha \bar z^{\bar \beta} \,,\;\;
\hat c_\epsilon(y,z) = c^a_{\alpha \epsilon} y^a z^\alpha \,.
\eea
In this regime, the maximization of $\Sigma$ with respect to the angles $\chi$ and $\xi$ can be performed explicitly.
When $|d| \gg |c|$, only the terms quadratic in $d_{abc}$ matter, and we see that these are positive for 
heterotic models and negative for orientifold models. The maximum $\Sigma_{\rm max}$ is  
then obtained for $\chi=\pi/2$ and $\xi = 0$ in heterotic models and for $\chi = 0$ in 
orientifold models:
\bea
\Sigma_{\rm max} \b\simeq\b \left\{ 
\begin{array}{l}
\displaystyle{\max_{y}} \Big\{\textstyle{\sum_r} |\hat d_r(y)|^2 \Big\} \;\;(\text{heterotic}) \,, \smallskip\ \\
0 \;\;(\text{orientifold}) \,.\\
\end{array}
\right.
\label{Sigmamaxlarged}
\eea
When on the contrary $|c| \gg |d|$, only the terms that are quadratic in $c^a_{\alpha \beta}$ matter. 
The maximum $\Sigma_{\rm max}$ is then obtained both in heterotic and orientifold 
models for $\chi= \pi/2$ and $\xi = \arcsin \sqrt{\sum_\epsilon |\hat c_\epsilon|^2/
(2 \sum_\epsilon |\hat c_\epsilon|^2 + \sum_r (c^r)^2)}$,
and one finds:
\bea
\Sigma_{\rm max} = \max_{y,z} \bigg\{\frac {2 \big(\sum_\epsilon |\hat c_\epsilon(y,z)|^2\big)^2}
{2 \sum_\epsilon |\hat c_\epsilon(y,z)|^2 + \sum_r (c^r(z))^2} \bigg\} \,.
\label{Sigmamaxlargec}
\eea
Finally, when $|c| \sim |d|$ one finds two different extrema, which generalize those seen above and 
compete against each other. The values of  $\Sigma$ at 
these two extrema can be computed explicitly, although we do not report their expressions here,
and $\Sigma_{\rm max}$ is then given by the maximum of these two extrema. 

Let us next consider the case where all the parameters $d_{abc}$ and $c^a_{\alpha \beta}$ are small.
In such a situation one may keep only those terms in (\ref{Sigmacan}) that involve no power of $d_{abc}$ 
and $c^a_{\alpha \beta}$. This leads to the following expression:
\bea
\Sigma(\chi,\xi,y,z) \b\simeq\b \sin^4\! \chi \bigg[\Big(\!-\! \text{\small $\frac {2 \pm 2}3$} 
+ \text{\small $\frac {2 \mp 1}3$} \zeta(y) \Big)\cos^4\! \xi 
- \text{\small $\frac 43$} \cos^2\! \xi \sin^2\! \xi \bigg] \nn \\[0mm]
\b\;\b +\,\sin^2\! \chi \cos^2\! \chi \bigg[ \!-\! \text{\small $\frac 83$}\, \kappa(x,y) \sin^2\! \xi \bigg] \,,
\label{Sigmasmalldc}
\eea
in terms of the following functions of $x$ and $y$, which take values in the interval $[0,1]$:
\bea
\a\a \zeta(y) = |\delta_{ab} y^a y^b|^2 \,,\;\; 
\kappa(x,y) = \text{\small $\frac 14$} \delta_{ab} (x \bar y^{\bar a} \!+\! \bar x y^a)(x \bar y^{\bar b} \!+\! \bar x y^b)\,.
\eea
In this regime, the maximization of $\Sigma$ with respect to the angles $\chi$ and $\xi$ can again be performed explicitly. 
For heterotic models, all the three terms are semi-negative definite. The maximum is thus obtained for $\chi = 0$ and 
gives the value $0$. For orientifold models, the first term is instead semi-positive definite while the other two terms are 
as before negative-definite. The maximum is then obtained for $\chi = \pi/2$ and $\xi = 0$ and gives the value $\zeta(y)$.
Since $\max_y \zeta(y) = 1$ we then get:
\bea
\Sigma_{\rm max} \b\simeq\b \left\{ 
\begin{array}{l}
0 \;\;(\text{heterotic}) \,, \smallskip\ \\
1 \;\;(\text{orientifold}) \,.\\
\end{array}
\right.
\label{Sigmamaxsmalldc}
\eea

The above results for the asymptotic behavior of $\Sigma_{\rm max}$ can be schematically summarized 
in the following simple way. When $|d|,|c| \gg 1$, one can have two types of behaviors: if $|d| \gg |c|$ then 
$\Sigma_{\rm max} \sim |d|^2$ for heterotic models but $\Sigma_{\rm max} \sim 0$ for orientifold models, 
while if $|d| \ll |c|$ then $\Sigma_{\rm max} \sim 2/3\, |c|^2$ both for heterotic and orientifold models, and 
when $|d| \sim |c|$ there is a transition between these two behaviors. When $|d|,|c| \ll 1$, one finds instead 
the following behavior: $\Sigma_{\rm max} \simeq 0$ for heterotic models and $\Sigma_{\rm max} \simeq 1$ 
for orientifold models. In terms of $a$ and $b$, this implies in particular that 
\bea
m^2{\!\!}_{f \bar f}  \b\simeq\b \left\{ 
\begin{array}{l}
\Big[\!-\! 4\, \sin^4\! \theta + 4 \sin^2\! \theta\, \cos^2\! \theta + (1 \pm 1)\, a \cos^4\! \theta\Big]\, m^2_{3/2} 
\,,\;\; a \gg 1, b \ll a\,, \medskip\ \\
\Big[\!-\! 4\, \sin^4\! \theta + 4 \sin^2\! \theta\, \cos^2\! \theta + 2\, b \cos^4\! \theta\Big]\, m^2_{3/2} 
\,,\;\; b \gg 1, a \ll b \,, \medskip\ \\
\Big[\!-\! 4\, \sin^4\! \theta + 4 \sin^2\! \theta\, \cos^2\! \theta + \displaystyle{\text{\small $\frac 32$}} 
(1 \mp 1) \cos^4\! \theta\Big]\, m^2_{3/2} \,,\;\; a \simeq -1, b \ll 1 \,.
\end{array} \label{mffcon}
\right.
\eea

\section{Soft masses and flavor universality}
\setcounter{equation}{0}

As a second application of the results derived in the previous sections, let us consider the condition for 
the flavor universality of soft supersymmetry breaking terms. This is controlled by the structure 
of soft scalar masses and depends on the holomorphic bisectional curvature of the scalar manifold 
along a given visible sector direction $v^I$ and the Goldstino direction $f^I$. More precisely, assuming 
again for simplicity a negligibly small cosmological constant, these masses are given by
\be
m^2_{v \bar v} = 3 \Big(R(v,f) + \text{\small $\frac 13$} \Big) \, m_{3/2}^2 \,,
\ee
where the holomorphic bisectional curvature $R(v,f)$ is defined as
\be
R(v,f) = - R_{I \bar J K \bar L} v^I \bar v^{\bar J}\! 
f^{\hspace{-1pt} K} \hspace{-3pt} \bar f^{\bar L}\,,
\ee
and the vectors $v^I$ and $f^I$ are subject to the following constraints:
\be
g_{I \bar J} v^I \bar v^{\bar J} = 1 \;,\;\; g_{I \bar J} f^I \! \bar f^{\bar J} = 1 \;,\;\;
g_{I \bar J} v^I \hspace{-1pt} \bar f^{\bar J} = 0 \,.
\ee
The condition of flavor universality is that $m^2{\!\!}_{v \bar v}$ be independent of $v^I$. 
A particularly simple and appealing first step in this direction could be to require that 
$m^2{\!\!}_{v \bar v} = 0$ for every $v^I$, which implies $R(v,f) = -1/3$. 
In the presence of a positive cosmological constant $V$ parametrized by $\gamma = V/(3\,m_{3/2}^2)$ 
this condition becomes $R(v,f) = -1/3 \, (1 + \gamma)^{-1}$. 
The effect of vector multiplets is instead discussed for example in \cite{FKZ,DV,S}.

In the class of models that we considered, the visible sector containing the standard particles 
must consist of a subset of the matter fields $\Phi^\alpha$, while the hidden sector can involve 
the dilaton $S$ and a subset of the K\"ahler moduli and matter fields $Z^i=\Phi^\alpha,T^A$. 
We thus have $v^S = 0$, $v^A = 0$, $v^\alpha \neq 0$, $f^S\neq 0$, $f^A \neq 0$ and $f^\alpha \neq 0$. 
As before, it is convenient to introduce an angle $\theta$ and write $f^S = \sin \theta\, g^S$ and 
$f^i = \cos \theta\, h^i$, where now $|g| = 1$ and $|h|=1$. 
We will again imagine that the Goldstino direction can a priori be arbitrary, as in \cite{KL,BIM},
and shall not discuss the possibilities offered by specific effects like classical fluxes or 
non-perturbative quantum corrections (see however \cite{CGGPS,DGM} for some recent studies 
on this applying to the minimal situation studied in this paper).
Noticing that the bisectional curvature 
of the fixed coset manifold $SU(1,1)/U(1)$ describing the dilaton is trivially $R(v,g) = 0$, and writing
the bisectional curvature of the generic no-scale manifold ${\cal M}_{Y,N}$ describing the K\"ahler
moduli and matter fields as $R(v,h) = -1/3 + \Xi(v,h)$, one can then write $R(v,f)$ in the following form:
\bea
R(v,f) = 0 \cdot \sin^2\! \theta + \Big(\!-\! \text{\small $\frac 13$} + \Xi(v,h) \Big) \cos^2\! \theta \,.
\eea
The soft scalar masses are correspondingly written as:
\bea
m^2_{v \bar v} = \Big[\sin^2\! \theta + 3\, \Xi(v,h) \cos^2\! \theta\Big] m^2_{3/2} \,.
\label{mvvcan}
\eea
The quantity $\Xi(v,h)$ can be non-zero only if the no-scale manifold ${\cal M}_{Y,N}$ differs from the 
minimal possibility $SU(1,1+n)/(U(1) \times SU(n))$. It measures the amount by which the bisectional 
curvature deviates from the critical value $-1/3$, and controls therefore the possibility of making 
$m^2_{v\bar v} \neq 0$ even when $\theta = 0$. A quite explicit but still general expression
for it can be derived by using the general properties of the geometry of no-scale manifolds derived in 
section 2, with $Y$ homogeneous of degree three in $J^A$ and $N^A$ function of $\Phi^\alpha \bar \Phi^\beta$,
under the simplifying assumption that the matter fields take vanishing expectation values.
Using the same short-hand notation as in the previous section, in which at the considered point the moduli 
index $A$ is split into the values $0$ corresponding to the direction parallel to $k^A$ and the values $a$ 
corresponding to the directions orthogonal to $k^A$, one finds:
\bea
\Xi(v,h) \b=\b P_{\alpha \bar \beta a \bar b}\, v^\alpha \bar v^{\bar \beta} h^a \bar h^{\bar b} 
+ Q_{\alpha \bar \beta \gamma \bar \delta} \, v^\alpha \bar v^{\bar \beta} h^\gamma \bar h^{\bar \delta} 
+ \text{\small $\frac 1{\sqrt{3}}$} \Gamma_{a \alpha \bar \beta} v^\alpha \bar v^{\bar \beta} 
\big(h^0 \bar h^{\bar a} + \bar h^{\bar 0} h^a \big) \,,
\eea
where:
\bea
\a\a P_{\alpha \bar \beta a \bar b} = \text{\small $\frac 13$} g_{\alpha \bar \beta} g_{a \bar b} 
- R_{\alpha \bar \beta a \bar b} \,, \label{P} \\
\a\a Q_{\alpha \bar \beta \gamma \bar \delta} = \text{\small $\frac 13$} 
\big(g_{\alpha \bar \beta} g_{\gamma \bar \delta} + g_{\alpha \bar \delta} g_{\gamma \bar \beta} \big)
- R_{\alpha \bar \beta \gamma \bar \delta} \label{Q} \,.
\eea
The explicit form of the normalization conditions is:
\be
g_{\alpha \bar \beta} v^\alpha \bar v^{\bar \beta} = 1 \,,\;\; 
|h^0|^2 + g_{a \bar b} h^a \bar h^{\bar b} + g_{\alpha \bar \beta} h^\alpha \bar h^{\bar \beta} = 1 \,,\;\;
g_{\alpha \bar \beta} v^\alpha \bar h^{\bar \beta} = 0\,.
\label{normgen}
\ee
Moreover, invariance under the visible sector gauge symmetries clearly implies that
\be
c^a_{\alpha \beta} v^\alpha \bar h^{\bar \beta} = 0 \,.
\ee

We now want to evaluate more explicitly the quantity $\Xi$ in the specific cases of Calabi-Yau 
string models of the heterotic and orientifold types, where the Riemann tensor and the Christoffel 
connection are parametrized in terms of some numbers $d_{ABC}$ and $c^A_{\alpha \beta}$.
To do so, it is again convenient to go to the canonical frame defined in sections 3 and 4. In this way, 
one can use the simple characterization of the geometry derived in section 5, and after a straightforward 
computation one finds that the quantities (\ref{P}) and (\ref{Q}) reduce to the following combinations of 
the quantities (\ref{b}) and (\ref{beta}):
\be
P_{\alpha \bar \beta a \bar b} = b^{ab}_{\alpha\beta} + \beta^{ab}_{\alpha\beta} - \frac 12 d_{abr} c^r_{\alpha\beta} \,,\;\;
Q_{\alpha \bar \beta \gamma \bar \delta} = - \big(c^a_{\alpha\beta} c^a_{\gamma\delta} 
+ c^a_{\alpha \delta} c^a_{\gamma\beta}\big) \,.
\ee
One then finds:
\bea
\Xi(v,h) \b=\b \Big(b^{ab}_{\alpha\beta} + \beta^{ab}_{\alpha\beta} - \frac 12 d_{abr} c^r_{\alpha\beta}\Big)
v^\alpha \bar v^{\bar \beta} h^a \bar h^{\bar b} 
-\, c^a_{\alpha\beta} c^a_{\gamma\delta} v^\alpha \bar v^{\bar \beta} h^\gamma \bar h^{\bar \delta} \nn \\ 
\b\;\b -\, \frac 1{\sqrt{3}} c^a_{\alpha\beta} v^\alpha \bar v^{\bar \beta} \big(h^0 \bar h^{\bar a} + \bar h^{\bar 0} h^a \big) \,.
\label{Xican}
\eea
We see that the structure of $\Xi(v,h)$ is absolutely identical in heterotic and orientifold models. 
The soft scalar masses correspondingly take the same form as derived in \cite{AS2} 
in both types of models. We further notice that $\Xi(h)$ has a very simple dependence on $h^0$.
This results again in two distinct behaviors for directions $h^i$ that are parallel and orthogonal to $k^i$. 
In the parallel direction with $h^0 =1$ and $h^a,h^\alpha = 0$, one finds a trivially vanishing $\Xi(v,h)$. 
In the orthogonal directions with $h^0 = 0$ and $h^a, h^\alpha \neq 0$, one instead finds a generically 
non-trivial $\Xi(v,h)$. Notice also that in hybrid directions where $h^a = 0$ and $h^0, h^\alpha \neq 0$, 
one finds again a vanishing $\Xi(v,h)$ if the further constraints $c^a_{\alpha \beta} h^\alpha \bar h^{\bar \beta} = 0$ 
hold true. In such a situation, the soft scalar masses would then become flavor universal:
\be
m^2{\!\!}_{v \bar v} = \sin^2\! \theta\, m^2_{3/2} \;\;\text{if}\;\; h^a = c^a_{\alpha \beta} h^\alpha \bar h^{\bar \beta} = 0 \,.
\label{mvvspecial}
\ee

\subsection{Possibility of mild sequestering}

We have just seen that one can achieve the critical value $\Xi(v,h) = 0$ in a rather simple and quite generic way by 
requiring the Goldstino direction $h^i$ to be such that $h^a = 0$ and imposing that $h^\alpha$ satisfies the further 
constraints $c^a_{\alpha\beta} h^\alpha \bar h^{\bar \beta}= 0$. Such constraints always admit at least 
one solution, which is $h^\alpha = 0$. This corresponds to taking $h^i$ parallel to $k^i$, which has indeed been 
shown to always yield $\Xi(v,h) = 0$. Under favorable circumstances, there may however also exist more general 
solutions with $h^\alpha \neq 0$. Whenever they arise, these correspond to a more general choice for $h^i$, 
which also yields $\Xi(v,h) = 0$ but in a potentially more flexible way. One may then try to investigate when such 
particular directions exist and whether it is possible to force the Goldstino direction to align along them as a result 
of a global symmetry, thereby realizing the idea of mild sequestering proposed in \cite{KMS} (see also \cite{SS}). 
It was however shown in \cite{AS2} that this is possible only whenever the matrices $c^a_{\alpha\beta}$ span a 
Lie algebra and $d_{abc}$ are the symmetric symbol of this algebra. This leads to the conclusion that such a 
mechanism is really natural only in models where the scalar manifold is a coset, and much less natural in models 
where the scalar manifold is generic.

\section{Conclusions}
\setcounter{equation}{0}

In this work, we have presented a general study of the geometry of no-scale K\"ahler manifolds.
We first derived a simple and novel general formula given by (\ref{Riemnoscale}) for the curvature tensor 
of a completely generic no-scale K\"ahler manifold, in the parametrization that naturally emerges 
in string models with some numbers of moduli $T^A$ and matter fields $\Phi^\alpha$, as a function 
of the metric and the third and fourth derivatives of $e^{-K}$. This result resembles very 
much the expression for the curvature tensor of special K\"ahler manifolds in special coordinates, and 
displays some peculiar properties. Most importantly, we showed that at every point of such a no-scale 
K\"ahler manifold there exists a special direction along which the sectional curvature has a universal 
critical value. We then studied in more detail the two classes of no-scale manifolds emerging from heterotic 
and orientifold string models based on a generic Calabi-Yau internal manifold with a generic gauge
bundle over it, characterized by some intersection numbers $d_{ABC}$ and some matrices 
$c^A\hspace{-6.5pt} \raisebox{-2pt}{\scriptsize $\alpha \beta$}$.
We restricted for simplicity to points where only the moduli fields and not the matter fields have non-vanishing 
values, and introduced a canonical parametrization at such a point, where the special direction of critical curvature 
is aligned with one of the moduli fields $T^0$, while the other orthogonal directions are associated 
to the other moduli fields $T^a$ and the matter fields $\Phi^\alpha$. We were then able to derive two 
very simple and similar expressions for the Riemann tensor in these two classes of no-scale manifolds, which 
are given by eqs.~(\ref{Riemstring1})--(\ref{Riemstring4}) as functions of the non-trivial components 
$d_{abc}$ and $c^a\hspace{-5pt} \raisebox{-2.5pt}{\scriptsize $\alpha \beta$}$ in the canonical frame defined at the 
reference point under consideration. We then gave a completely algebraic characterization of the conditions 
under which such manifolds become symmetric cosets, showing that the deviations from such a 
situation are essentially controlled by two combinations of parameters, called $a_{abcd}$ and 
$b^{ab}\hspace{-9.5pt} \raisebox{-3pt}{\scriptsize $\alpha \beta$}$ and defined by (\ref{a}) and (\ref{b}).
This allowed us to argue that while in the case of one modulus field and any number of matter fields one unavoidably 
gets a maximally symmetric manifold, and in the case of two moduli fields and zero matter fields one finds 
a disconnected one-parameter family of models separated by a unique possible coset manifold, in all other cases 
one obtains a connected multi-parameter family of models where possible coset manifolds represent isolated points.
We then observed that the no-scale manifolds arising in heterotic and orientifold models display a kind of duality, 
in the sense that the associated Riemann tensors differ only by the sign of the contribution depending on $a_{abcd}$, 
while all the remaining terms and in particular those depending on $b^{ab}\hspace{-9.5pt} \raisebox{-3pt}{\scriptsize $\alpha \beta$}$
have the same sign. As a result, the heterotic and orientifold no-scale manifolds coincide if $a_{abcd}$ vanishes,
while $b^{ab}\hspace{-9.5pt} \raisebox{-3pt}{\scriptsize $\alpha \beta$}$ may still be arbitrary, so that one may or 
may not get a coset manifold. 

As an application of the results that we derived for the geometry of no-scale K\"ahler manifolds, we studied the general 
structure of those scalar masses that are entirely controlled by supersymmetry breaking splitting effects, in string 
models where the universal dilaton sector and a generic no-scale sector involving an arbitrary number 
of K\"ahler moduli and matter fields are included. We used for this the general form that the K\"ahler potential must take 
in such a situation and assumed that a completely generic superpotential may arise and trigger supersymmetry breaking
in way involving all the above fields. 
As a first application, we studied the average sGoldstino square mass $m^2{\!\!}_{f \bar f}$ in the hidden sector of 
superfields taking non-vanishing expectation values, defined by the Goldstino direction $f^i$ of supersymmetry breaking. 
This direction has components $f^S$ and $f^A,f^\alpha$ in the dilaton and no-scale sectors, with a relative magnitude
that is weighted by an angle $\theta$.
We derived an explicit expression for $m^2{\!\!}_{f \bar f}$ in the canonical frame, given by eqs.~(\ref{mffcan}) and (\ref{Sigmacan}),
and showed that it is essentially controlled by the quantities 
$a_{abcd}$ and $b^{ab}\hspace{-9.5pt} \raisebox{-3pt}{\scriptsize $\alpha \beta$}$ that parametrize the deviations 
of the geometry from a coset situation. More precisely, what matter are the extremal values $(\pm a)_{\rm up}$ and 
$b_{\rm up}$ that can be achieved for their contractions $\pm a(\tilde y)$ and $b(\tilde y, \tilde z)$ along real normalized 
directions $\tilde y^a$ and $\tilde z^\alpha$ in the subspaces of the non-minimal moduli $T^a$ and the matter 
fields $\Phi^\alpha$, the two signs applying respectively to heterotic and orientifold models. We first showed that a
qualitative necessary condition for being able to achieve even for vanishing $\theta$ a positive $m^2{\!\!}_{f \bar f}$, which 
is necessary and sufficient for the existence of a metastable supersymmetry breaking vacuum if one allows the 
superpotential to be tuned, is that at least one of the two quantities $(\pm a)_{\rm up}$ and $b_{\rm up}$ be positive. 
We then also derived an upper 
bound on the absolute magnitude of $m^2{\!\!}_{f \bar f}$, which also represents an upper bound on the mass of the lightest 
particle in the hidden sector, given by (\ref{mffcon}). In particular, this formula shows that when the effects of moduli 
fields dominate one finds at best $m^2{\!\!}_{f \bar f} \simeq 2\, a\, m^2_{3/2}$ for $a \gg 1$ in heterotic models and 
$m^2{\!\!}_{f \bar f} \simeq 16/5\, m^2_{3/2}$ when $a \simeq - 1$ in orientifold models, while when the effects of matter 
fields dominate one can achieve $m^2{\!\!}_{f \bar f} \simeq 2\,b\, m^2_{3/2}$ for $b \gg 1$ in both models.
We finally argued that $m^2{\!\!}_{f \bar f}$ can generically be made hierarchically larger than $m^2_{3/2}$ by suitable 
choosing the vacuum point to make $a$ or $b$ large. More precisely, with a single modulus and any number of matter 
fields $m^2{\!\!}_{f \bar f}$ is bounded, and with any number of moduli and zero matter 
fields it can be arbitrarily large in heterotic models and is bounded in orientifold models, but in any other situation one 
can get an arbitrarily large result both for heterotic and orientifold models, except for the isolated cases corresponding 
to coset manifolds. As a second application we studied the soft scalar square masses $m^2{\!\!}_{v \bar v}$ in 
the visible sector of superfields taking vanishing expectation values, defined by an arbitrary direction $v^\alpha$ in flavor space. 
We presented a simple general expression for $m^2{\!\!}_{v \bar v}$  in the canonical frame, given by eqs.~(\ref{mvvcan}) and (\ref{Xican}), and emphasized that it is identical in form for heterotic and orientifold models. We first investigated the conditions 
under which $m^2{\!\!}_{v \bar v}$ can be flavor universal, as required by phenomenological considerations. We then showed 
that $m^2{\!\!}_{v \bar v}$ may be forced to be flavor universal by suitably orienting the Goldstino direction and that this mild 
sequestering mechanism may be implemented by postulating the existence of some approximate global symmetries 
in the hidden sector.

To conclude, let us remark that the general and model-independent results we derived for $m^2{\!\!}_{f \bar f}$
and $m^2{\!\!}_{v \bar v}$ also display interesting correlations. Most importantly, we see that when the Goldstino direction 
satisfies the constraints $f^a= c^a\hspace{-5pt} \raisebox{-2.5pt}{\scriptsize $\alpha \beta$} f^\alpha \bar f^{\bar \beta} = 0$,
one interestingly finds that $m^2{\!\!}_{f \bar f}$ is given by the bounded and sign-indefinite result (\ref{mffspecial}) 
and $m^2{\!\!}_{v \bar v}$ is given by the flavor-universal and positive result (\ref{mvvspecial}). This seems to suggest 
that there is no cheap way of simultaneously achieving a large $m^2{\!\!}_{f \bar f}$ and a flavor universal $m^2{\!\!}_{v \bar v}$.

\vskip 20pt

\noindent
{\bf \Large Acknowledgements}

\vskip 10pt

\noindent
This work was supported by the Swiss National Science Foundation.
We thank M.~Trigiante for useful discussions and K.~Sinha for interesting correspondence.

\end{document}